\documentclass[12pt]{article}

\usepackage{tikz}
\usepackage{url}
\usepackage{amsmath}
\usepackage{amsfonts}
\usepackage{amssymb}
\usepackage{footmisc}
\usepackage{longtable}
\usepackage{amsmath,amsfonts,amssymb,amsthm}
\usepackage{enumerate}
\newcommand{\vs}{{\beta}}
\newcommand{\bs}{\rho}

\newcommand{\E}{\mathbb{E}}
\newcommand{\mwr}{\texttt{mwr}}
\usepackage{soul}
\definecolor{sangre}{rgb}{0.6,0.18,0.19}
\definecolor{dullmagenta}{rgb}{0.4,0,0.4}
\definecolor{darkblue}{rgb}{0,0,0.6}
\usepackage{natbib}

\ifx\pdfoutput\undefined
\usepackage[backref=none,hypertex,colorlinks,urlcolor = darkblue,citecolor = sangre,linkcolor=blue]{hyperref}
\else
\usepackage[backref=none,hypertexnames=false,colorlinks,urlcolor = darkblue,citecolor =  sangre,linkcolor=blue]{hyperref}
\fi

\usepackage{relsize,etoolbox}
\usepackage{bbm}
\usepackage{color, colortbl}
\definecolor{Gray}{gray}{0.95}
\usepackage{booktabs}
\usepackage{pdflscape}
\usepackage{amsfonts}
\usepackage{graphicx,epstopdf}
\usepackage{subcaption}

\usepackage{multirow}
\usepackage{epstopdf}
\usepackage{adjustbox}
\usepackage[titletoc,title]{appendix}

\usepackage[capitalize]{cleveref}
\definecolor{lavander}{cmyk}{0,0.48,0,0}
\definecolor{violet}{cmyk}{0.79,0.88,0,0}
\definecolor{burntorange}{cmyk}{0,0.52,1,0}

\def\lav{lavander!90}
\def\oran{orange!30}

\tikzstyle{peers}=[draw,circle,violet,bottom color=\lav,
                  top color= white, text=violet,minimum width=10pt]
\tikzstyle{superpeers}=[draw,circle,burntorange, left color=\oran,
                       text=violet,minimum width=20pt]
\tikzstyle{legendsp}=[rectangle, draw, burntorange, rounded corners,
                     thin,bottom color=\oran, top color=white,
                     text=burntorange, minimum width=2.5cm]
\tikzstyle{legendp}=[rectangle, draw, violet, rounded corners, thin,
                     bottom color=\lav, top color= white,
                     text= violet, minimum width= 2.5cm]
\tikzstyle{legend_general}=[rectangle, rounded corners, thin,
                           burntorange, fill= white, draw, text=violet,
                           minimum width=2.5cm, minimum height=0.8cm]

\usepackage[most]{tcolorbox}
 \tcbset{colback=white}
\theoremstyle{plain}

\usepackage{setspace}
\usepackage[margin=1in]{geometry}

\usepackage{caption}
\usepackage{graphicx}
\usepackage{graphics}
\usepackage{color}
\usepackage{pstricks}
\usepackage{sgame}
\usepackage{egameps}
\usepackage{amsthm}
\usepackage{mathrsfs}
\usepackage{appendix}
\makeatletter
\pgfdeclareshape{datastore}{
  \inheritsavedanchors[from=rectangle]
  \inheritanchorborder[from=rectangle]
  \inheritanchor[from=rectangle]{center}
  \inheritanchor[from=rectangle]{base}
  \inheritanchor[from=rectangle]{north}
  \inheritanchor[from=rectangle]{north east}
  \inheritanchor[from=rectangle]{east}
  \inheritanchor[from=rectangle]{south east}
  \inheritanchor[from=rectangle]{south}
  \inheritanchor[from=rectangle]{south west}
  \inheritanchor[from=rectangle]{west}
  \inheritanchor[from=rectangle]{north west}
  \backgroundpath{
    \southwest \pgf@xa=\pgf@x \pgf@ya=\pgf@y
    \northeast \pgf@xb=\pgf@x \pgf@yb=\pgf@y
    \pgfpathmoveto{\pgfpoint{\pgf@xa}{\pgf@ya}}
    \pgfpathlineto{\pgfpoint{\pgf@xb}{\pgf@ya}}
    \pgfpathmoveto{\pgfpoint{\pgf@xa}{\pgf@yb}}
    \pgfpathlineto{\pgfpoint{\pgf@xb}{\pgf@yb}}
 }
}
\theoremstyle{definition}

\newtheorem{lemma}{Lemma}
\newcommand{\distas}[1]{\mathbin{\overset{#1}{\kern\z@\sim}}}%
\newsavebox{\mybox}\newsavebox{\mysim}
\newcommand{\distras}[1]{%
  \savebox{\mybox}{\hbox{\kern3pt$\scriptstyle#1$\kern3pt}}%
  \savebox{\mysim}{\hbox{$\sim$}}%
  \mathbin{\overset{#1}{\kern\z@\resizebox{\wd\mybox}{\ht\mysim}{$\sim$}}}%
 }
\usepackage{xcolor}
\usepackage[linesnumbered,ruled,vlined]{algorithm2e}

\usepackage{soul}
\begin{document}

\title{Auctioning Annuities\thanks{\footnotesize We are thankful to Zachary Bethune, Manudeep Bhuller, David Byrne, Liran Einav, Leora Friedberg, Gast\'on Illanes, Elena Krasnokutskaya, Fabian Lange, Lee M. Lockwood, Fernando Luco, John Pepper and Chris Yung     and  for their suggestions and feedbacks. We also thank the seminar and conference participants and discussants at SEA 2019, APIOC 2019, 2020 ES NAWM, NYU Stern, ES World Congress 2020, INFORMS 2020, IIOC 2021 and DC IO Day 2021. Fajnzylber and Willington acknowledge financial support from CONICYT/ANID Chile, Fondecyt Project Number 1181960. The project described received funding from the TIAA Institute. The findings and conclusions expressed are those of the authors and do not necessarily represent official views of the TIAA Institute, TIAA, or the Inter-American Development Bank.}
}
\author{Gaurab Aryal\footnote{University of Virginia, e-mail: {aryalg@virginia.edu}},   Eduardo Fajnzylber\footnote{Inter-American Development Bank, e-mail: {eduardofa@iadb.org}},\\ Maria F. Gabrielli\footnote{Universidad del Desarrollo and CONICET, e-mail: {fgabrielli@udd.cl}}, and Manuel  Willington\footnote{Universidad del Desarrollo, e-mail: {mwillington@udd.cl}}} 
\date{\today}

\maketitle
\begin{abstract}
We propose and estimate a model of demand \emph{and} supply of annuities. To this end, we use rich data from Chile, where annuities are bought and sold in a private market via a two-stage process: first-price auctions followed by bargaining. We model firms with private information about costs and retirees with different mortalities and preferences for bequests and firms' risk ratings. We find substantial costs and preference heterogeneity, and because there are many firms, the market performs well. 
Counterfactuals show that simplifying the current mechanism with English auctions and ``shutting down" risk ratings increase pensions, but only for high-savers.\\

\end{abstract}

\section{Introduction}
In many countries, policymakers are ``rethinking" social retirement programs \citep{Feldstein2005, MitchellShea2016} and are considering relying more on market-based schemes to provide better terms for retirement products.\footnote{Market-based systems have been used successfully by the Centers for Medicare and Medicaid Services (CMS) to reduce expenses \citep{GAO2016} and health exchanges under the \emph{Affordable Care Act of 2010}.} One such product is an annuity contract, which provides insurance against financial volatility and the risk of outliving one's savings \citep{Yaari1965, BrownMitcheellPoterbaWarshawsky2001, DavidoffBrownDiamond2005}.\footnote{An annuity is an insurance contract where in exchange for a lump-sum payment an insurer promises a stream of payments to the annuitant until her death. It may also include payments for beneficiaries after her death.} A key question then is how to ``design" markets for annuities, and what are the associated distributional consequences. For instance, allowing greater individual control can improve retirees' welfare, but if retirees also use non-price attributes (e.g., firms' risk ratings) in their decision, it may lower effective competition and their welfare.  

The answer depends on demand \emph{and} supply factors and how they interact to determine pensions and welfare.
While many research papers focus on the demand \citep[e.g.,][]{EinavFinkelsteinSchrimpf2010, IllanesPadi2019}, we know relatively little about the strategic supply side of annuities.\footnote{Also see \cite{HastingsHortacsuSyverson2017}, who develop and estimate a model of fund manager choice by workers to study the role of the sales force in Mexico's privatized social security system.} 
However, estimating both of them poses several challenges. For one, retirees differ in their savings and expected longevity, but they may also have different preferences for bequests and non-price attributes. Second, life insurance companies may differ in terms of their expectations about a retiree's longevity. Third, they may also have different (opportunity) costs of promising survival-contingent payments based on their investment portfolios; these costs are likely to be firms' private information.

In this paper, we propose and estimate a model of demand and strategic supply of annuities that incorporates preference heterogeneity and asymmetric information about firms' costs. 
We provide new empirical evidence on the performance of a private market for annuities and quantify how differences in preferences and costs across retirees and firms affect individual pensions and welfare. 
To this end, we use a rich administrative dataset from the annuity market in Chile, structured as first-price auctions followed by bargaining between a retiree and several firms. We use our estimates to compute the welfare costs of asymmetric information and gains --in terms of pensions and welfare-- of using more standard and simpler selling mechanisms like English auctions.\footnote{This counterfactual exercise is motivated by an ongoing policy debate in Chile about ways to increase pensions. Chilean antitrust authority believes that pensions are low, in part, because the current pricing mechanism is complex and retirees have a poor understanding of the role of firms' risk ratings \citep{FNE2018}. So there is a proposal in the parliament to use simpler auctions, where the highest pension offer wins.} 

 Chile provides an ideal setting to study these questions as it has a relatively ``thick" annuity market, with an annuitization rate close to 60\%.\footnote{This is in sharp contrast with the \emph{annuity puzzle} \citep{Lockwood2012} literature that has attempted to reconcile the relatively low annuitization rate observed elsewhere. One plausible explanation is that in Chile, the alternative to an annuity is not a lump-sum (as in other cases) but a programmed withdrawal (PW). Under PW, payments are determined every year as if they were an actuarially fair annuity. As a consequence, the stream of pension payments decreases until the funds are eventually exhausted.}
 Moreover, in Chile, all firms must use the same centralized electronic platform, with the acronym SCOMP, and they can sell only \emph{fixed annuities} with constant (real) payments that are simpler for retirees to understand --and for us to model-- than \emph{variable annuities} \citep{NYT2016}.\footnote{Contracts may differ in terms of guaranteed periods and deferral periods. As we explain below, these features determine the \emph{expected present discounted value} of payments that retirees or their heirs enjoy.} Thus, Chile is at the frontier for understanding competitive annuities market, and our empirical findings may suggest lessons to countries that either have or are considering a market-based system.\footnote{For example, in the U.S., the \emph{SECURE Act of 2019} incentivizes businesses and communities to band together to offer annuities. However, the law is silent about how to structure such markets.}  

Our modeling assumptions and empirical choices are motivated by the Chilean retirement process and our dataset to achieve our goal. We have information on all retirees who used SCOMP between January 2007 and December 2017.\footnote{Every worker with pension savings above an exogenous threshold \emph{must} use the SCOMP at retirement.} We observe everything firms observe on every retiree before making their first-round offers. In particular, we observe retirees' demographic characteristics, savings, the list of all participating firms and their pension offers for different types of annuities, and retirees' final choices. We also observe the date of death of retirees who passed away within the observation window. Using this mortality data, we estimate a continuous-time duration model and predict each retiree's expected longevity. 


The timing of the ``game'' is as follows. First, the retiree requests offers from insurers on several types of annuities through SCOMP. Then, a subset of active firms (with potentially different risk ratings) ``enter" and simultaneously offer different types of annuities. We posit that the retiree with a commonly known expected longevity calculates the \emph{expected present discounted utility} associated with each of these offers, and she either chooses the outside option of programmed withdrawal (PW) or chooses one of the offers or chooses to bargain with firms for better offers on one type of annuity. Most retirees in our sample opt to bargain. We model the second-stage bargaining process as a multi-attribute oral ascending auction \citep{BulowKlemperer1996}, where the chosen firm offers the second-largest expected present discounted utility that any of the losing firms can offer.\footnote{Following \cite{KrasnokutskayaSongTang2020}, we refer to such auctions as multi-attribute auctions, where unlike scoring-auctions \citep{AskerCantillon2008} some bidder attributes (risk ratings) are exogenous.}

 Demand for annuities also depends on bequest-preferences \citep{KopczukLupton2007, Lockwood2012, EinavFinkelsteinSchrimpf2010, IllanesPadi2019}. 
However, some retirees may prefer annuities without bequests because they cannot afford bequests or leave bequests through a non-annuity channel, e.g., houses, or they do not have any heirs. We model bequest-preferences as a random coefficient with an unknown mass at zero to capture this heterogeneity across retirees. 

Besides pensions and bequests, retirees may also care about firms' risk ratings, although they may also differ in their understanding of the role of this attribute. 	
To capture this uncertainty, we use the framework in \cite{MatejkaMcKay2015} and model retirees as \emph{rationally-inattentive} decision-makers \citep{Sims1998} who have to process costly information to learn about the value of risk ratings. If the cost varies across retirees, their posterior beliefs will also vary, and they act ``as if" they value risk ratings differently.\footnote{The bad risk rating can be a proxy for a firm's financial insolvency, although that risk in Chile is negligible. Although we focus only on risk ratings, there may be other sources of ``friction." For instance, in the context of the U.S., \cite{BrownKapteynLuttmerMitchell2017} suggests that retirees are unable to compare different types of annuities. In Chile, however, SCOMP simplifies the annuitization process and provides information necessary for retirees to compare different pensions (see, for example, Figure \ref{fig:scomp_picture}). Given this, here we focus on risk-rating and use rational inattention to capture parsimoniously the fact that there is widespread uncertainty about the usefulness of risk rating, yet firms may overstate (during the bargaining process) its relevance.}

To model annuitization costs, we note that retirees differ in savings and demographic characteristics, so firms' annuitization costs are retiree-specific.
These costs depend on firms' expectations about retiree's life expectancy and, more importantly, on their financial positions, their asset-management liabilities, and investment opportunities \citep{RochaThorburn2007}. Thus, we model firms as symmetric bidders competing for retirees' savings. 

Then, to identify preferences and costs, we exploit {contract choices}, {equilibrium} conditions, and exogenous variations in demographic characteristics and savings.
To identify the distribution of bequest preferences, we use data from first-round offers {for different contracts} that have different ``price gradients" for bequests and ``trace" how bequest choices vary with these gradients. Finally, to identify the information processing cost, we use the fact that demand elasticity with respect to pensions is inversely proportional to the cost of processing information. 
Furthermore, to identify preferences for firms' risk ratings and the cost distributions, we focus only on the bargaining stage and adapt the identification strategies from random coefficient models \citep{HoderleinKlemelaMammen2010} and English auctions to our setting of multi-attribute oral ascending auctions.  

Our first set of results suggest that retirees' information-processing costs decrease with their savings. 
Second, those who use sales agents or directly contact insurance companies have higher information-processing costs and value risk ratings more than others. 
Thus, while everyone starts with a prior that gives a positive value to risk ratings, those with low information-processing costs revise the value towards zero. 
Third, while more than 40\% of annuitants do not value bequest, there is a significant variation among the rest. 

Fourth, we also find substantial heterogeneity in annuitization costs across retirees and their savings.  
If we had considered only the average annuitization costs, we would miss this heterogeneity, as the mean cost is similar across savings quintiles. 
In particular, we find that the top $40\%$ of savers have a higher probability ($14\%$ versus $6\%$) of having annuitization costs below actuarially fair costs than low savers. Furthermore, the left tails of these cost distributions are ``fatter" for high savers and, as there are almost always 13 to 15 active firms, these tails are key in determining equilibrium pensions.

Finally, we (i) determine the welfare costs of asymmetric information and find that it is non-negligible for only the top 40\% of savers; and (ii) evaluate the effect of replacing the current pricing mechanism with English auctions while ``shutting down'' the role of risk ratings. A priori, this simplification should increase ``effective" competition and pensions for low savers since they value risk rating more. We find, however, that only the top $40\%$ of savers benefit from the new scheme because of the fatter tails of the cost distributions. These pension increases, however, do not translate into significant utility gains as a result of diminishing marginal utilities.\footnote{Throughout the paper, we follow the literature, e.g., \cite{MitchellPoterbaWarshawskyBrown1999, EinavFinkelsteinSchrimpf2010} and use homogenous Bernoulli utility with constant relative risk-aversion coefficient of $3$.}

Altogether these results shed light on the main factors that affect annuity markets. 
For instance, we find that a simpler pricing mechanism would benefit all retirees, but the gains are distributed asymmetrically across retirees depending on firms' costs of annuitizing their savings. Nevertheless, we can also say that the Chilean market is functioning well, primarily because of the competitive supply with many active firms.

We end by observing that these results depend on our model assumptions. In particular, we assume that firms know retirees' preferences and their expected longevity for model tractability.\footnote{If retirees have private information about their longevity, then modeling the supply side becomes hard because we have to model firms with multi-dimensional signals: one on the common value component (longevity) and the other on the private value component (annuitization costs); see, for example, \cite{GoereeOfferman2002}. However, extending their model to our setting is beyond the scope of our paper.} 
This assumption enabled us to develop a novel supply-side analysis, but it rules out the adverse selection, a phenomenon that has been found in previous research; see, e.g., \cite{FinkelsteinPoterba2002, EinavFinkelsteinSchrimpf2010} for the U.K and \cite{FajnzylberWillington2019, IllanesPadi2019} for Chile. However, this assumption may not be as strong in our setting as it may appear.
First, \cite{IllanesPadi2019} observe that the selection on longevity risk is mitigated by the selection on ``non-cost dimensions of preferences," such as the bequest preference. 
Second, unlike in the U.K., the alternative to an annuity in Chile is PW, and PW is more similar to an annuity than cash withdrawal. So it is reasonable to expect a milder selection.

Nonetheless, to minimize selection bias, we restrict attention only to annuitants and never compare their annuitization costs or bequest preferences to those choosing PW. We also consider selected policy interventions that are less likely to affect the ``selection margin.'' In that sense, we do not attempt to assess the welfare effects of allowing a lump sum withdrawal.

 We proceed as follows. Sections \ref{section:institution} and \ref{section:data} introduce annuities and data, respectively. Sections \ref{section:model} and \ref{section:identification} present the model and identification strategy, respectively. Sections \ref{section:estimation} and \ref{section:counterfactual} present our results, and in section \ref{section:conclusion}, we conclude. Other details are in the Appendices A-F. 

\section{Background on Annuities\label{section:institution}}
\subsection{Institutional Detail}
The Chilean pension system went through a major reform in the early 1980s, when it transitioned from a \emph{pay-as-you-go} system to a system of fully funded capitalization in individual accounts run by private pension funds (henceforth, AFPs). 
Under the new system, workers must contribute $10\%$ of their monthly earnings, up to a predetermined maximum (which in 2018 was U.S. \$2,319), into savings accounts managed by the AFPs.\footnote{This maximum, and annuities in general, are expressed in \emph{Unidades de Fomento} (UF), a unit of account used in Chile that closely follows the CPI. On December 31st, 2017, 1 UF was approximately equivalent to U.S. \$43.38.\label{footnote:UF} Throughout the paper, pensions, and savings are expressed in U.S. dollars unless stated otherwise.} The total balance in this account at the time of retirement minus any eligible withdrawals is what we refer to as savings. 
The normal retirement age is 60 years for women and 65 years for men, but workers with sufficient funds may retire earlier. 
We focus only on retirees who have savings above the regulated threshold, and by law, \emph{have to} participate in SCOMP. 

On the supply side, the government heavily regulates the life insurance industry, and the current regulatory framework recognizes that the main risks associated with annuities are the risks of longevity and reinvestment. 
To deal with the longevity risk, firms who want to sell annuities have to maintain technical reserves.
The government also regularly assesses the risk of reinvestment via the Asset Sufficiency Test established in 2007. 
Under this regulation, if and when there are ``insufficient" asset flows, an insurance company must establish additional technical reserves.

Bankruptcies among life insurance companies are rare in Chile. Nonetheless, the government guarantees every retiree pensions up to 100\% of the basic solidarity pension \citep{Fajnzylber2018}, and 75\% of the excess pension over this amount, up to 45 UF (see Footnote \ref{footnote:UF}).

\subsection{Different Types of Annuities}
Retirees have three main choices: programmed withdrawal (PW), immediate annuity (IA), and deferred annuity (DA). Under PW, savings remain under the AFP management and are paid back to the retiree following a predetermined benefit schedule that steadily decreases over time. After death, any remaining funds are used to finance survivorship pensions or become part of the retiree's inheritance in the absence of eligible beneficiaries. 
So PW benefits are exposed to financial volatility and do not provide longevity insurance.

Under IA and DA, a retirees' savings are transferred to an insurance company of her choosing, and that firm provides an inflation-indexed monthly pension to her and her surviving beneficiaries. 
Under DA, pensions are contracted for a future date (usually between one and three years), and in the meantime, the retiree is allowed to receive a temporary benefit that can be as high as twice the deferred annuity amount. 

Thus, the main trade-off between an annuity and a PW is that an annuity provides insurance against longevity risk and financial risk, whereas under a PW, a retiree can bequeath all remaining funds in case of early death. Moreover, while annuitization is an irreversible decision, a retiree who chooses a PW can switch to an annuity at a later date.

Annuities may also include a particular coverage clause called the guaranteed period (GP). For example, if an annuity contract includes 10-years guaranteed period, then either the retiree or her eligible beneficiaries get the full pension for ten years. After the guaranteed period, the contract reverts to the standard annuity.

\begin{figure}[h!]
\caption{\bf Example: Benefit Schedules, by Annuity Type}
\centering
\includegraphics[scale=0.5]{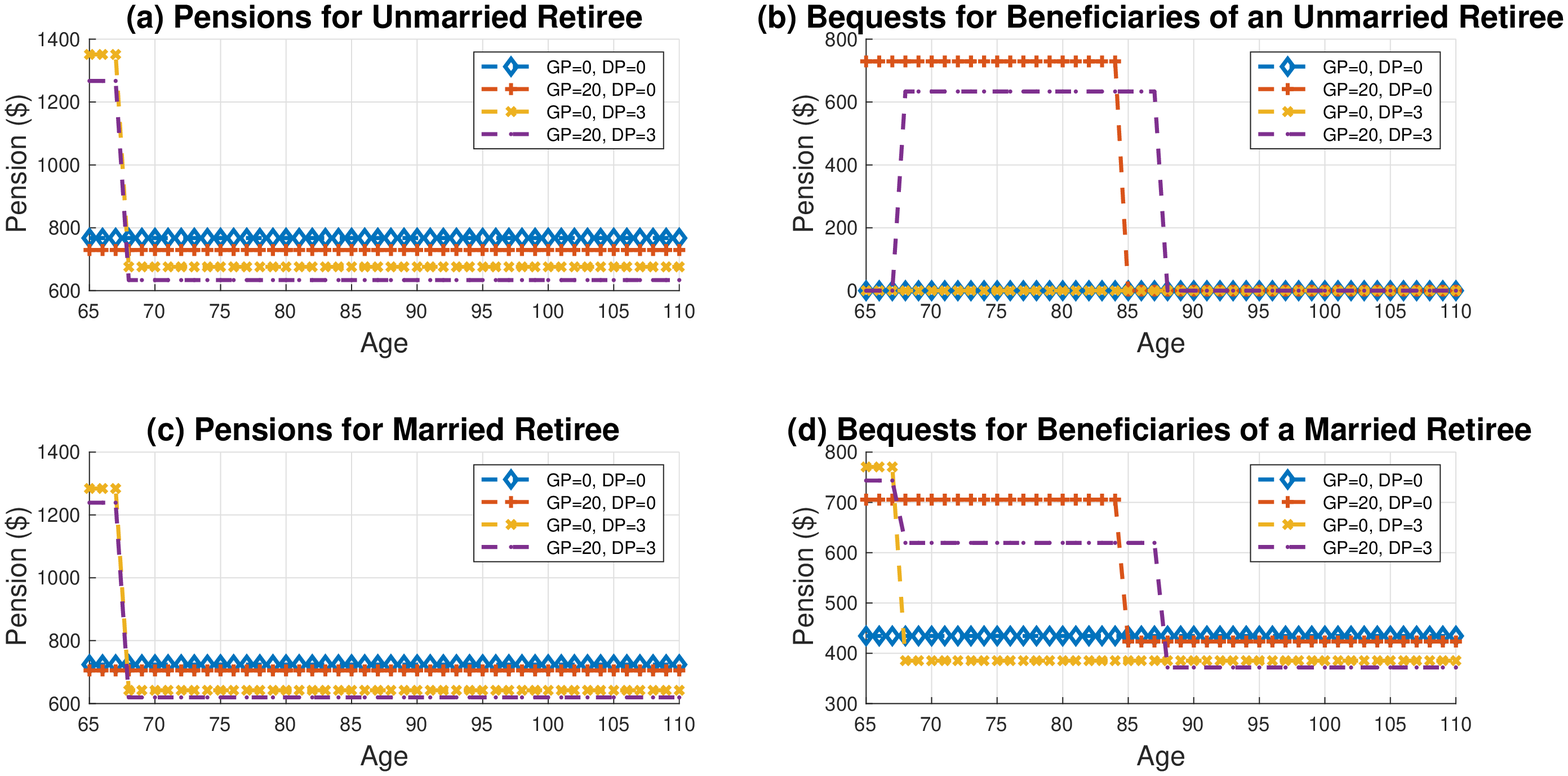}
\label{fig:benefit}
\caption*{\footnotesize {\bf Note:} The figure shows the survival-contingent benefit schedules for retirees and their beneficiaries. For this example, we take a 65 years old male retiree with savings of U.S. \$200,000 in 2020 and use the official \emph{mortality table}. Subfigures (a) and (b) show the pension and bequest schedules, respectively, for four types of annuities if the retiree is unmarried. Subfigures (c) and (d) show the pension and benefit schedules if he is married. GP stands for a guaranteed period (years), and DP stands for a deferred period (years). }
\end{figure}

{\bf Example.} For an illustration of how benefits change with different types of annuities and marital status, consider a retiree who is 65 years old male with a savings of US\$200,000 and is retiring in 2020. 
If the retiree is unmarried and chooses an annuity with GP=0 and DP=0; the pension is constant until death (blue `$\diamond$' in Figure \ref{fig:benefit}-(a)), and after that the heirs get nothing (blue `$\diamond$' in Figure \ref{fig:benefit}-(b)). However, if the retiree chooses an annuity with GP=20, then the retiree gets a lower pension when alive (red `$+$' vs. blue `$\diamond$' in Figure \ref{fig:benefit}-(a)). If the retiree dies within 20 years of retirement, then the designated beneficiaries get a positive amount (red `$+$' in Figure \ref{fig:benefit}-(b)) until 20 years after retirement, and after that, they get nothing. If the retiree was married, then even with GP=0 and DP=0 (blue `$\diamond$' in Figure \ref{fig:benefit}-(c)), the beneficiaries (in this case, the surviving spouse) will get a positive amount (blue `$\diamond$' in Figure \ref{fig:benefit}-(d)) after the retiree dies. 	

\subsection{Steps in Buying an Annuity \label{section:retirement}}
The process begins when a worker communicates her decision to retire to her AFP. 
We assume that she is then exogenously matched with one of four intermediaries or ``channels" --AFP, direct contact with an insurance company, sales agent, and independent advisor--who can help her choose an annuity and a firm. 
Out of the four channels, only the first two are free.\footnote{We treat this match as exogenous. For more on this see Appendix \ref{section:channels}.} 
 Retirees must also disclose information on all eligible beneficiaries, i.e., spouses and children, and the AFP generates a \emph{Balance Certificate} with information on her savings, hers, and her legal beneficiaries' demographic characteristics. Then the process is as follows: 
\begin{enumerate}
\item A retiree requests offers for up to 12 different types of annuities and PW. Then her \emph{Balance Certificate} is shared with all the insurance companies in the system, who then have eight business days to make an offer for those products. 

\item These offers are collected and collated by SCOMP and presented to the retiree as a \emph{Certificate of Quotes}. 
The certificate is in the form of tables, one for each type of annuity, sorted in terms of pensions and includes firms' risk ratings; see Figure \ref{fig:scomp_picture}.
\item The retiree has five options: (i) postpone retirement; (ii) file a new request for quotes; (iii) choose PW; (iv) accept one of the offers; or (v) negotiate with companies by requesting second-round offers. In the latter case, firms cannot offer below their initial-round offers, and the individual can always fall back to any first-round offer.
\end{enumerate}

\section{Data\label{section:data}}
We have administrative data from SCOMP on all annuities purchased in Chile from January 2007 to December 2017.
We observe everyone who used SCOMP to buy an annuity or choose PW during this period. 
As mentioned before, we observe everything about a retiree that all life insurance companies observe before making their participation (``entry") decisions and their first-round offers.  
We also observe all first-round offers each retiree received, their final choices, and whether they chose in the second round. 
\subsection{Retirees}
We focus on individuals without eligible children and who retire within ten years of the ``normal retirement age" (NRA), which is 60 years for women and 65 years for men.
In total, we observe 238,891 retirees, with an almost even split between PW, immediate annuities, and deferred annuities; see Table \ref{freq_tipo_pension}. Less than 1\% of retirees choose a combination of annuity \emph{with} PW, and so we exclude them, leaving a total of 238,548 retirees.
\begin{table}[th!]\centering 
\caption{\label{freq_tipo_pension} 
\textbf{Share of Pension Products}}
\scalebox{0.85}{\begin{tabular} {@{} l r r @{}}  \toprule
{\bf Product}& {\bf Obs.} & \% \\
\toprule
PW&   78,161&       32.7\\
Immediate annuity&   87,115&       36.4\\
Deferred annuity&   73,272&       30.6\\
Annuity with PW&      343&        0.9\\
\hline
Full Sample&  238,891&      100\\
\bottomrule
\end{tabular}}
\caption*{\footnotesize {\bf Note.} The table shows the distribution of retirees across different annuity products. We restrict ourselves to annuities with either 0, 10, 15 or 20 years of guaranteed periods or at most 3 years of deferment.}

\end{table}

In Table \ref{table:gendermarried}, we present the sample distribution by retirees' marital status, gender, and retirement age. 
Around 56\% retire at their NRA and 79\% retiree at or at most within three years after NRA (rows 2 and 3). Married men are half of all retirees.
Retirees also vary in terms of their savings; see Table \ref{tab:balance_by_age_group}. 
The average savings in our sample are \$112,471, while the median savings are \$74,515 with an inter-quartile range of \$85,907. 
 Savings are higher for men and for those who retire before NRA. 
\begin{table}[htbp!]\centering
\caption{\label{table:gendermarried} 
\textbf{Age Distribution, by Gender and Marital Status}}
\scalebox{0.85}{\begin{tabular}{cccccc}
\toprule
\textbf{Retiring Age} & 
\textbf{S-F} &\textbf{M-F} & \textbf{S-M} & \textbf{M-M} &\textbf{Total} \\  \toprule
Before NRA&    1,871&    1,771&    4,714&   22,142&   30,498\\
At NRA&   20,789&   22,475&   17,114&   72,572&  132,950\\
Within 3 years after NRA&   14,470&   16,797&    4,447&   19,086&   54,800\\
At least 4 years after NRA&    6,900&    6,715&    1,251&    5,434&   20,300\\\hline 
Full Sample&   44,030&   47,758&   27,526&  119,234&  238,548\\ \bottomrule 
 \end{tabular}}
\caption*{\footnotesize {\bf Note.} The table displays the distribution of retirees, by their marital status, gender and their retirement ages. 
Thus the first two columns `S-F' and `M-F' refer, respectively, to single female and married female, and so on. NRA is the `normal retirement age,' which is 60 years for a female and 65 years for a male. }
\end{table}


For each type of annuity, a retiree receives an average of 10.6 offers in the first round. Moreover, the number of offers increases with savings. 
For example, retirees with savings at the $75^{th}$ percentile of our sample get an average of $12.4$ offers, and those at the $25^{th}$ percentile get an average of 7.8 offers. 
It is reasonable to assume that retirees with higher savings are more lucrative for the firms, and therefore more companies are willing to annuitize their savings. 
If, however, those with higher savings also live longer than those with lower savings, then it means that annuitizing higher savings is costlier for firms.
\begin{table}[ht!]\centering \caption{\bf Savings, by Retirement Age and Gender \label{tab:balance_by_age_group}}
\scalebox{1}{\begin{tabular}{l c c c c c}\toprule
 & {\bf Mean} & {\bf Median} & {\bf P25} & {\bf P75}& {\bf N} \\
 \toprule
 {\bf Retiring Age}& &  & & &  \\
Before NRA&185,660&129,637&73,104&245,857&30,498 \\
At NRA&89,907&60,023&41,521&103,680&132,950 \\
Within 3 years after NRA&115,666&87,126&54,353&135,562&54,800 \\
At least 4 years after NRA&141,673&101,594&58,815&168,202&20,300 \\
 \cline{2-6}
Full Sample&112,471&74,515&46,449&132,356&238,548 \\
\midrule
{\bf Gender}&&&&& \\
Female&97,308&81,180&51,817&121,633&91,788 \\
Male&121,955&69,372&43,818&147,184&146,760 \\
\cline{2-6}
Full Sample&112,471&74,515&46,449&132,356&238,548 \\

\bottomrule
\end{tabular}}
\caption*{\footnotesize {\bf Note:} Summary statistics of savings, in U.S. dollars, by retiree's age at retirement, and by retiree's gender.}
\end{table}

Offered pensions vary across life insurance companies and across retirees; see Table \ref{tab:pension_gender_type}.
For an IA, retirees get an average offer of \$570, and for DA, the average offer is \$446. 
On average, women get an offer of \$479 for IA and \$412 for DA, while for men, they are \$631 and \$473, respectively. These features are consistent with men having more savings \emph{and} shorter life expectancy than women. See the estimated longevity in Table \ref{table:mediantime}.

\begin{table}[th!]
\begin{center}\caption{\bf Monthly Pension Offers, by Annuity Type and Gender\label{tab:pension_gender_type}}
\scalebox{0.85}{\begin{tabular}{cccc|ccccc}
\toprule
%

&&   &  & {\bf Savings} & {\bf Savings}& {\bf Savings}& {\bf Savings}& {\bf Savings} \\

{\bf Annuity Type}  &{\bf Gender} 	& {\bf Mean}& {\bf Median}     & {\bf Q1} & {\bf Q2} & {\bf Q3}& {\bf Q4}& {\bf Q5} \\
\toprule
Immediate &Female     &479 	&414 	& 202	& 288		& 385		& 510		& 857	 \\
    		 &Male 	    &631    &435         &200	& 269                & 372	        & 585	& 1329	 \\
\cline{2-9}
	&	Full Sample&570&423&201	& 278& 378& 556& 1152	 \\
	\cline{2-9}
Deferred &Female&412&374&190 & 258& 349& 463& 714\\
&Male&473&356&187& 241& 331&529& 1019 \\\cline{2-9}

 &Full Sample&446&365&189& 248& 339& 500& 882 \\

\bottomrule

 \end{tabular}}
\caption*{\footnotesize {\bf Note:} Summary of average monthly pensions (in U.S. dollars) from offers received in the first round. 
}\end{center}
\end{table}

To capture this variation in pensions, we allow firms to have different annuitization costs, also known as the \emph{Unitary Necessary Capital} (henceforth UNC).\footnote{The per-dollar annuitization cost is known as the \emph{Unitary Necessary Capital} (UNC). It captures the cost of making a survival-contingent stream of payments. In particular, UNC is the expected amount of dollars required to finance a stream of payments of one dollar until the retiree's death and any proportional obligations to her surviving relatives, if any. For example, if the UNC of a firm is 200, it means that the firm's expected cost to provide a pension of \$100 is \$20,000. } 
Firms' UNCs can be different for the same retiree, based on firms' assessments of expected longevity and the opportunity costs of committing a survival-contingent dollar payment.
We posit that only firms know their annuitization costs and use asymmetric information to model firms' decisions.

In Table \ref{goes2round_by_accept_type} we show the choices made by retirees across different stages. 
Most retirees who choose PW do that in the first round (98.1\%), and most retirees (86.9\%) who choose annuity do that in the second round.
As we can see, only 2,979 retirees opt during the second round but choose an annuity quote from the first round.  
\begin{table}[htbp]\centering
\caption{\label{goes2round_by_accept_type} 
\textbf{ Number of Retirees who choose in First- or Second-Round }}
\scalebox{0.85}{\begin{tabular}{ccccc}
\toprule
{\bf Round/Choice} &{\bf PW} &{\bf $1^{st}$ round} &{\bf $2^{nd}$ round} &{\bf Total} \\  \toprule
$1^{st}$ round&   76,690&   18,001&        0&   94,691\\
$2^{nd}$ round&    1,471&    2,979&  139,407&  143,857\\
\midrule
Total&   78,161&   20,980&  139,407&  238,548\\\bottomrule 
\end{tabular}}
\caption*{\footnotesize {\bf Note.} Round refers to whether retirees chose in the first- or in the second-round.} 
\end{table}


\subsection{Firms\label{section:firmdata}}
In our sample, we observe 20 unique life insurance companies with different risk ratings, which are generally constant throughout our sample period. Most companies have high (at least AA) risk ratings. 
Table \ref{table:ratings} shows the distribution of risk ratings. 
For our empirical analysis, we treat these ratings as exogenous and group them into three categories: 3 for the highest risk rating (AA+), 2 for intermediate-risk ratings (from AA to A), and 1 for the rest (below A). Not all 20 firms are active at all times, and not all participate in every retiree-auction.
On average, 11 firms participate in every particular annuity type auction. 

We define \emph{potential entrants} (for each retiree auction) as the set of active firms that participated in at least one retiree-auction in the same month. In our sample, retirees have either 13, 14, or 15 potential entrants.
\begin{table}[ht!]
 \centering
 \caption{\bf Risk Ratings}\label{table:ratings}
\scalebox{1}{ \begin{tabular}{lccc}
\toprule
 {\bf Rating} & {\bf Frequency} & \% & {\bf Cumulative \%} \\
 \toprule
 AA+ & 155 & 24.64 & 24.64 \\
 AA & 245 & 38.95 & 63.59 \\
 AA- & 171 & 27.19 & 90.78 \\
 A+ & 2 & 0.32 & 91.1 \\
 A & 15 & 2.38 & 93.48 \\
 BBB+ & 1 & 0.16 & 93.64 \\
 BBB & 6 & 0.95 & 94.59 \\
 BBB- & 15 & 2.38 & 96.98 \\
 BB+ & 19 & 3.02 & 100 \\
 \midrule
 Total & 629 & 100 & \\
 \bottomrule
 \end{tabular}}%
 \caption*{\footnotesize {\bf Note:} The table shows the distribution of quarterly credit ratings from 2007-2018.}
\end{table}%

The participation rate --the ratio of the number of actual bidders to the number of potential bidders-- varies across retirees: it ranges from as low as 0.08 to as high as 1, with mean and median rates of 0.73 and 0.78, respectively, and a standard deviation of 0.18.\footnote{ Using a Poisson regression of the number of participating firms on the retiree characteristics, we find that one standard deviation increase in savings, which is approximately \$87,000, is associated with roughly one more entrant. Moreover, women have 0.61 more participating companies than men, while sales agents and advisors are associated with approximately 0.19 fewer participants than the other two channels.}  
Thus, it is reasonable to assume that a firm's decision to participate depends on its financial position when a retiree initiates the process.\footnote{ We tested this selection by estimating a Heckman selection model with the number of potential bidders as the excluded variable and found strong evidence of negative selection among firms.}  
To capture this selection, in our empirical application, we follow \cite{Samuelson1985} to model firms' entry decisions, which posits that firms observe their retiree-auction-specific annuitization cost before entry.  
 
Next, we implement a simple diagnostic analysis to determine if it is reasonable to assume that firms have symmetric cost distributions. 
In particular, for each firm, we use a linear regression model to residualize the pensions it offers. Then, we compare the distributions of the residuals, one for each firm. If these errors have similar distributions, then we say that firms are symmetric. Otherwise, we model firms as asymmetric firms.

Let the pension-rate $\texttt{Pension-Rate}_{i,j}=P_{ij}/S_i$ be the ratio of pension offered by $j$ to retiree $i$ to $i$'s savings. Then we estimate by ordinary least squares the following model:
\begin{eqnarray}\label{eq:symmetry}
		 \texttt{Pension-Rate}_{i,j}&=& \texttt{constant} +\beta_1 \times UNC_{i}+\beta_{2} \times \texttt{Age}_{i}+\beta_{3}\times \texttt{Gender}_{i}\notag\\
		&&+ \beta_{4}\times\texttt{Marital Status}_{i}+\beta_{5}\times \texttt{Spouse's Age}_{i}\notag\\
		&&+\beta_{6} \times \texttt{Guaranteed Months}_{i}+\beta_7 \times\texttt{Potential Bidders}_{i}+\varrho_{i,j},\qquad
\end{eqnarray}
where all the regressors are observed in the data except for $i$'s Unitary Necessary Capital ($UNC_i$), which measures the expected discounted number of months retiree $i$ is expected to live after retirement. So a retiree who expects to live longer will have a larger $UNC_i$, and will be costlier for firms to annuitize. 
For the same retiree, UNCs across firms can be different, based on firms' assessments of expected longevity and the opportunity costs of committing a survival-contingent dollar payment. 
We formally define $UNC_i$ later when we present the model, and in the Appendix \ref{section:Gompertz} we provide details on how to estimate the mortality distributions and then use the estimated distributions to determine $UNC_i$.\footnote{In Figure \ref{fig:UNCi} we display the histograms and scatter plots of monthly pension per annuitized dollar and the $UNC_i$'s of all the retirees that the firms made offers to in the first stage.}  

\begin{figure}[t!]\caption{\bf Distributions of Homogenized Pension Rates, by Firms\label{fig:symmetry}}
 \begin{center}
 \includegraphics[scale=0.3]{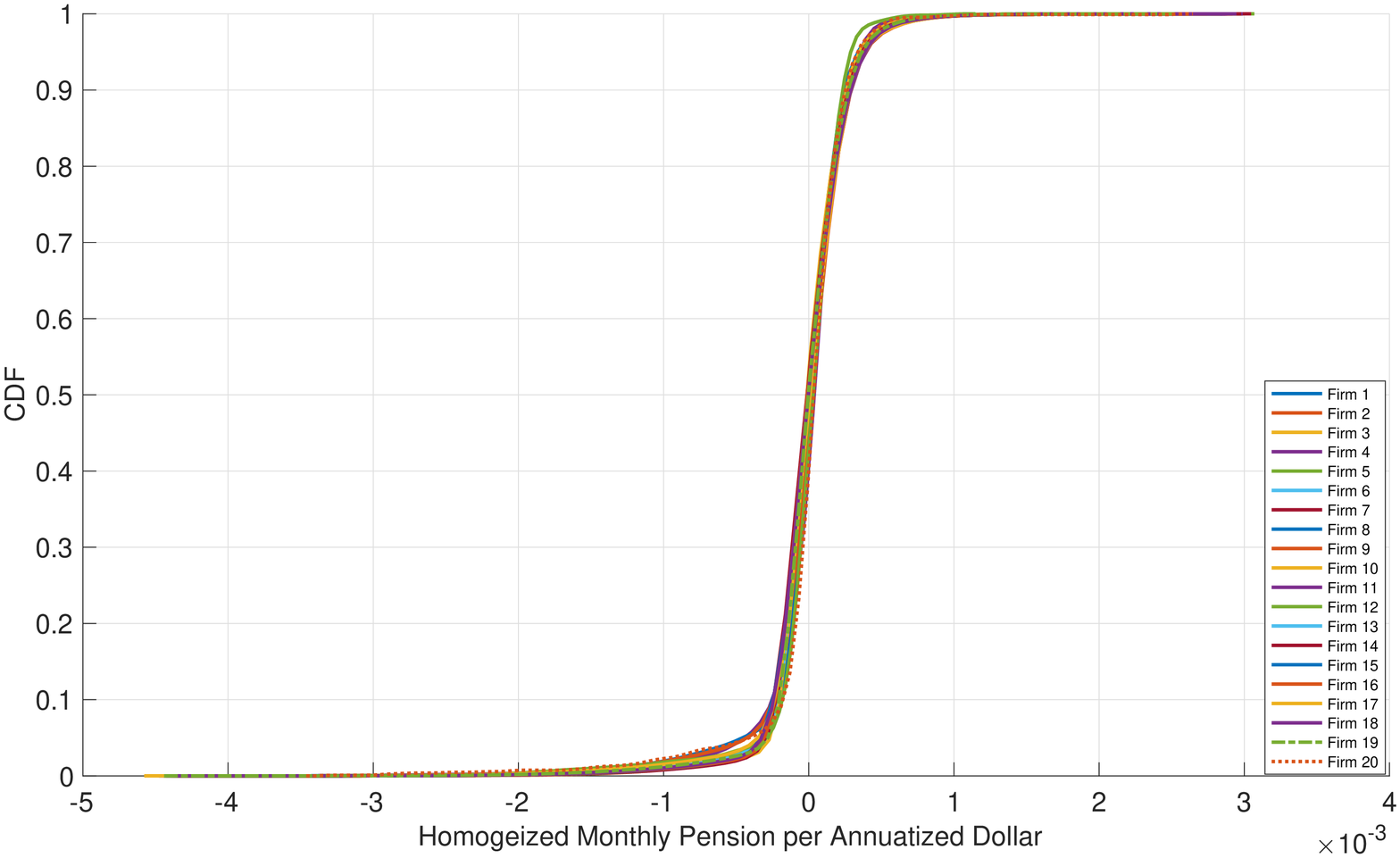}  
\caption*{\footnotesize {\bf Note:} Kernel density estimates of the distributions of $\hat{\varrho}_{ij}$ from Equation (\ref{eq:symmetry}), one for each of the 20 firms. }
\end{center}
\end{figure}

We estimate (\ref{eq:symmetry}) separately for each firm $j=1,\ldots, 20$, and predict the firm-retiree-specific residual $\hat{\varrho}_{i,j}$ for retiree $i$.
    Then in Figure \ref{fig:symmetry}, we display the Kernel density estimate of $\hat{\varrho}_{i,j}$ for $j=1,\ldots, 20$.
As we can see, all the distributions are almost identical, which suggests that it is reasonable to assume that firms have symmetric cost distribution but allow the distribution to vary with retirees' savings. 

\section{Model\label{section:model}}
In this section, we introduce our model. 
We consider the decision problem facing a retiree who uses SCOMP to choose a company to annuitize her savings to model the demand. To model the utility from an annuity, we closely follow the extant literature on annuities, with a modification to model retirees as rationally inattentive decision-makers with respect to their preferences for risk ratings. 

To model the imperfect competition among insurance companies, we use multi-attribute auctions with selective entry. First, the firms decide to participate, and conditional on participating, they compete in a first-price auction with independent private values. In the second stage, firms compete in multilateral bargaining with one-sided asymmetric information, which we model as an oral ascending auction. We implicitly assume that while bargaining firms also learn about retirees' preferences for bequests and risk-rating.

\subsection{Demand\label{section:demand}}
Here we consider the problem faced by an annuitant $i$, who has already decided which annuity product to choose (e.g., an immediate annuity with 0 guaranteed period) and considers between $J_i$ firms that have decided to participate in the auction for $i$'s savings $S_i$. 
The retiree will choose the firm that provides her the highest indirect utility.
 
We assume that the utility from an annuity consists of three parts: (i) the \emph{expected present discounted utility} from the pensions that the retiree consumes while alive, (ii) the utility she gets from leaving a bequest to her kin, and (iii) her preference for firm's risk rating. 
We model retirees as rationally inattentive decision-makers, but we begin without rational inattention for ease of exposition.  
 
Let $(\theta_i, \beta_i)$ denote $i$'s preferences for bequest and risk rating, respectively. 
And conditional on savings $S$, let $(\theta, \beta)$ be distributed independently and identically across retirees as ${F}_{\theta|S}(\cdot|S)\times F_{\beta|S}(\cdot|S)$ on $[0, \overline{\theta}]\times [\underline{\beta}, \overline{\beta}]$. 
Retirees may not afford bequests or may have additional wealth outside our sample for their heirs. 
In both cases, retirees may behave ``as if" they do not care about bequest. To capture this ``mass at zero" we allow the distribution ${F}_{\theta|S}(\cdot|S)$ to have a mass point at $\theta=0$, and the mass can depend on savings $S$. Let $\zeta(S)\in(0,1)$ be the probability that the retiree has $\theta_i=0$, and we define $F_{\theta|S}(\cdot|S)= \zeta(S) + (1-\zeta(S))\times \tilde{F}_{\theta|S}(\cdot|S)$ where, $\tilde{F}_{\theta|S}(\cdot|S)$ is the continuous distribution on $(0, \overline{\theta}]$ with $\lim_{t\rightarrow0^+}\tilde{F}_{\theta|S}(t|S)=0$. 

Let $P_{ij}$ denote the pension offered by firm $j$ to retiree $i$. 
For any annuity, with a pension offer $P_{ij}$, there is a corresponding bequest $B_{ij}$, which also depends on $i$'s expected mortality and the mortality of her beneficiaries. 
Whenever possible, we suppress the dependence of the bequest $B_{ij}$ on the pension $P_{ij}$.
Let $i$'s indirect utility at retirement from choosing $(P_{ij}, B_{ij})$ from firm $j$ with risk rating $Z_{j}\in\{1,2,3\}$ be  
\begin{eqnarray}
{U}_{ij} = \underbrace{\mathfrak{U}(P_{ij}, B_{ij};\theta_i)}_{\texttt{expected present discounted utility}}+ \underbrace{\beta_{i}\times {Z}_{j}}_{\texttt{utility from risk rating}} - \underbrace{\mathfrak{U}_{0i}(S_i).}_{\texttt{utility from programmed withdrawal}}\label{eq:utility1}
\end{eqnarray}

Next, we explain each of the three functions on the right-hand side of (\ref{eq:utility1}). 
The expected present discounted $\mathfrak{U}(P_{ij}, B_{ij};\theta_i)$ is a function of the retiree's expected mortality and the utility she gets from pension and bequest. Although pension is constant, the associated bequest can vary with time, depending on the retiree's death and the contract features. 
To determine the expected present discounted utility, we implement the following steps. First, we estimate the mortality process using a continuous-time proportional hazard model under the assumption that mortality follows a Gompertz distribution conditional on retiree's demographics and savings. 

Second, to determine utility from pension and bequest, respectively, following \cite{MitchellPoterbaWarshawskyBrown1999} and \cite{EinavFinkelsteinSchrimpf2010}, we assume that retirees have homothetic preferences with CRRA utility such that utility from pension $P$ is $u(P) = \frac{P^{(1-\gamma)}}{ 1-\gamma}$, and the utility from bequest $B$ is $v(B) = \frac{B^{(1-\gamma)}}{ 1-\gamma}$, both with $\gamma=3$. Third, we determine the associated pension and (time-dependent) bequest for each annuity and determine the expected present discounted utility under the assumption that retirees consume all their pensions. The discount factor is the market rate of return.  

We explain these steps in detail in Appendix \ref{section:Gompertz}, but it suffices to know that we can re-write the expected present discounted utility $\mathfrak{U}(P_{ij}, B_{ij};\theta_i)=\rho_i(P_{ij})+ \theta_i \times b_i(P_{ij})$, where $\rho_i(P_{ij})$ and $b_i(P_{ij})$ are the expected present discounted utility from pension $P_{ij}$ and bequest $B_{ij}$, which is proportional to $P_{ij}$. 
Substituting $\mathfrak{U}(P_{ij}, B_{ij};\theta_i)=\rho_i(P_{ij})+ \theta_i \times b_i(P_{ij})$ in (\ref{eq:utility1}) we can write $i$'s indirect utility from annuity $P_{ij}$ as 
\begin{eqnarray}
{U}_{ij}&=& \rho_{i}(P_{ij})+\theta_i\times b_{i}(P_{ij})+{\beta_{i}\times{Z}_{ij}}-\mathfrak{U}_{0i}(S_i).\label{eq:utility}
\end{eqnarray}

Annuities are complex financial contracts, and although there is evidence that raises questions about retirees' ability to compare different types of annuities, \citep[e.g.,][]{BrownKapteynLuttmerMitchell2017}, unlike in the U.S., in Chile, they can compare different choices with relative ease. This is so because, as explained in Section \ref{section:institution}, there are only a few standardized contracts, and the retiree receives all quotes in an easy to compare format: For each quoted product, the retiree gets a list of all monthly pension offers sorted from highest to lowest (see Figure \ref{fig:scomp_picture}). Comparing different offers for the same product is immediate, and comparing different products (e.g., annuities with a different guaranteed period) is not too onerous.

Thus (\ref{eq:utility}) shows a trade-off between higher pensions and lower risk ratings, but as mentioned above, we assume that $i$ does not know her $\beta_i$, but only its distribution.
We follow \cite{MatejkaMcKay2015} and assume that before the retirement process begins, $i$ believes that $\vs_{i}\distras{i.i.d} F_{\vs}(\cdot)$ with support $[\underline{\vs}, \overline{\vs}]$, and if $i$ wants to learn her preference, she has to incur an information-processing cost, valued at $\alpha>0$ per unit of information, which can vary by savings, which in turn can correlate with financial literacy, e.g., \cite{BrownKapteynLuttmerMitchell2017}, and, more importantly, also with the entry channel. The channel may affect the cost of learning $\beta$ if, for example, the sales agent exaggerates the importance of risk rating.

So, $i$ has first to decide how much to spend learning about $\vs_{i}$, and after that make the decision.
Let $\sigma:[\underline{\vs}, \overline{\vs}]\times {\mathcal P} \rightarrow \Gamma:=\Delta([0,1]^{J+1})$ denote the strategy of a retiree with preference parameter $\vs$, with offered pensions ${\boldsymbol P}_i:=({P}_{i1},\ldots, {P}_{iJ})\in{\mathcal P}$.  
The strategy is a vector 
$
\sigma(\vs,{\boldsymbol P}_i)\equiv \left(\sigma_{1}(\vs,{\boldsymbol P}_i),\ldots, \sigma_{J}(\vs,{\boldsymbol P}_i), \sigma_{J+1}(\vs, {\boldsymbol P}_i)\right)$ 
of probabilities, where $\sigma_{j}(\vs,{\boldsymbol P}_i)=\Pr(\texttt{$i$ chooses $j$}|\vs, {\boldsymbol P}_i)\in[0,1]$.
For notational simplicity, we suppress the dependence of choice probabilities on the offers $({\boldsymbol P}_i)$.
Then, by adapting \cite{MatejkaMcKay2015}'s choice formula to two periods, the probability that $i$ chooses $j$ is given by 
\begin{eqnarray}
\sigma_{ij}({\boldsymbol P}_i)=\sigma_{j}(\vs_i,{\boldsymbol P}_i)=\left\{\begin{array}{cc}\frac{\exp\left(\log \sigma_{j}^{0}+\frac{U_{ij}}{\alpha}\right)}{\sum_{k=1}^{J}\exp\left(\log \sigma_{k}^{0}+\frac{U_{ik}}{\alpha}\right)+\exp\left(\frac{\E U_{i}}{\alpha}\right)}, & j=1,\ldots, J\\
&\\
\frac{\exp\left(\frac{\E U_{i}}{\alpha}\right)}{\sum_{k=1}^{J}\exp\left(\log \sigma_{k}^{0}+\frac{U_{ik}}{\alpha}\right)+\exp\left(\frac{\E U_{i}}{\alpha}\right)}, & j=J+1,\end{array}\right.\label{eq:prob}
\end{eqnarray}
where $\sigma_j^0:=\int_{\underline{\beta}}^{\overline{\beta}} \sigma_j(\beta)dF_\beta(\beta)$ is the unconditional probability of choosing firm $j$, and $\E U_i$ is $i$'s ex-ante expected utility from choosing to go to the second round. As we focus on the second round, to keep the notations simple, we do not give the exact expression for $\E U_i$.

Finally, with a slight abuse of notation, we use $\mathfrak{U}_{0i}(S_i)$ to denote the utility from PW under which, and as we mentioned in the introduction, payments are determined every year as if it is an actuarially fair annuity. For our purposes, however, it is not necessary to calculate $\mathfrak{U}_{0i}(S_i)$ because we focus only on the annuitants and the identification depends on the differences in utilities that $i$ gets from different firms, canceling out $\mathfrak{U}_{0i}(S_i)$. 

\subsection{Supply\label{section:supply}}
Next, we present the supply side, where $J$ firms compete for $i$'s savings $S_i$. To this end, we first introduce the annuitization cost and its distributions before discussing the competition. Here, we take the entry decision as given and characterize the equilibrium only for the last stage of the game, i.e., the bargaining stage. 
\subsubsection*{Annuitization Costs.} Companies have different $UNC$s for the same retiree, depending on the demographic characteristics and the firms' portfolio and their asset-liability position \citep{RochaThorburn2007}. 
Thus, if $j$ can annuitize $i$ at a lower cost than $j'$, then $j$ has an advantage over $j'$ because all else equal, $j$ can offer a higher pension. 

Let $UNC_j^R$ be $j$'s unitary necessary capital to finance a dollar pension for the retiree. 
Similarly, we must consider the costs related to the bequest, which may come from two sources: a guaranteed period, during which after the death of the retiree, the beneficiaries receive the full amount of the pension, and the compulsory survival benefit, according to which the spouse of the retiree receives after her death and after the guaranteed period is over, $60\%$ of the pension until (spouse's) death.\footnote{For the exact form, see the derivation of Equation (\ref{eq:w}) in the Appendix \ref{section:Gompertz}.} We denote by $UNC_j^{S,GP}$ and $UNC_j^{S}$ the present value of the cost of providing these two benefits, respectively. 
Then, putting these costs together, $j$'s expected cost of offering $P_{ij}$ is given by\footnote{AFP pays the retirees during the deferred periods. So we omit them from the firm's $UNC$.}
\begin{eqnarray}
C(P_{ij}):=P_{ij}\times (UNC_j^R+ 0.6 \times UNC_j^{S} + UNC_j^{S,GP})\equiv P_{ij}\times UNC_j. \label{eq:expcost}
\end{eqnarray}
We define $UNC_i$ as the unitary cost of a pension calculated with the mortality process we estimate and the retirees' discount rate, proxied by the average market return rate. 
For the same retiree $i$, firms' $UNC$s may differ from $UNC_i$ due to the differences in their (i) mortality estimates, (ii) investment opportunities, and (iii) expectations about future interest rates.
For these reasons, only firm $j$ likely knows its $UNC_j$. 
Moreover, the ratio of $UNC_j$ to $UNC_i$ captures $j$'s relative efficiency selling an annuity to $i$, and we call this ratio $r_{ji}\equiv \frac{UNC_j}{UNC_i}$, $j$'s relative cost of annuitizing a dollar. Working with $r$, we can compare costs across retirees who have different $UNC_i$'s.

We assume that firms are symmetric (see Figure \ref{fig:symmetry} in Section \ref{section:firmdata}) and the cost $r_{ji}$ is private information and is distributed independently and identically across companies as $W_{r|S}(\cdot|S)$, with density $w_{r|S}(\cdot|S)$ that is strictly positive everywhere in its support $[\underline{r}, \overline{r}]$. 
We allow the cost distribution to depend on $S$ to capture the fact that those who have higher savings tend to live longer (Table \ref{table:mediantime}) and therefore are costlier to annuitize. 

\subsubsection*{\bf First-Stage Bidding.} If we set aside the second round, and the multi-product nature of the first round, $j$'s net present expected profit from offering $P_{ij}$, to a retiree $i$ with $S_i$ is  
\begin{eqnarray}
\E{\Pi}_{ij}^I(P_{ij})&=&(S_i - P_{ij}\times UNC_j)) \times \Pr(\texttt{j is chosen by offering} \quad\!\! P_{ij}|{\bf P}_{i-j})\notag\\
&=&S_i \times {(1 - r_{ji} \times \bs_{i}^*(P_{ij}))}\times \sigma_{ij}({\bf P}_i),\label{eq:profit}
\end{eqnarray}
where $\bs_{i}^*(P_{ij})\equiv P_{ij}\times UNC_i/S_i$ and $\sigma_{ij}({\bf P}_i)$ is the probability that $i$ chooses $j$ given the vector of offers ${\bf P}_i$. 
With the second round, $j$'s ex-ante expected profit is given by
\begin{eqnarray}
S_{i} \times(1-r_{ji}\times \bs_{i}^*(P_{ij}))\times \sigma_{ij}({\bf P}_i) + \sigma_{iJ+1}({\bf P}_i)\times \E \Pi_{j}^{II}(\bs_i^*(\tilde{P}_{ij})|r_{ji},{\bf P}_i),
\label{eq:eprofit}
\end{eqnarray}
where $\tilde{P}_{ij}$ is $j$'s second-round offer and $\sigma_{iJ+1}({\bf P}_i)$ from (\ref{eq:prob}) is the probability that $i$ takes the bargaining option in the second round with expected profit given by $\E \Pi_{j}^{II}$. 

The two rounds are connected. First, more generous offers on the first round may lower the retiree's probability of going to the second round. Second, and more importantly, each firm's first-round offer is binding for the second round: A firm cannot make any second-round offer below its first-round one. 

Now, when we include the fact that $i$ might request offers from $A_i$ types of annuities, insurance companies have to solve a multi-product bidding problem. 
As mentioned in the timing assumptions, once $i$ receives all the offers $\{P_{ij}^a: a\in A_i, j\in J\}$, she chooses $a^*\in A_i$ and then chooses the firm.
Thus, with a slight abuse of notations, we can express the expected profit of a firm $j\in J$ from an auction where $i$ requests offers for $A_i$ types of annuities as $\E\Pi_{ij}:= \sum_{a\in A_i} \E\Pi_{ij}(a) \times \Pr(\texttt{i chooses $a$}|\{{\bf P}_{i}^b\}_{b\in A_i}; \theta_i).$

Thus, in the first round, when choosing $P_{ij}^a$, firm $j$ has to consider the competition from other firms for product $a$ and all other types of annuities in $A_i\backslash \{a\}$. It also has to consider competition from its offers $P_{ij}^b, b\in A_i, b\neq a$, which is the cannibalization consideration facing multi-product firms. Determining the equilibrium bidding strategies for the first round auction requires us first to determine the equilibrium in the bargaining phase. 

\subsubsection*{\bf Second-Stage Bargaining.} Next, we characterize the equilibrium in the second stage bargaining game among $J$ firms under the assumption that by the time the bargaining starts, the retiree knows which type of annuity to buy, and firms commonly learn about retiree's $(\theta_i, \beta_i)$. Thus, the winner is the firm that can offer the highest utility.  

We base the assumption that firms also learn about retirees' $(\beta, \theta)$ on the observation that there are many interactions between firms and retirees, so it is reasonable that firms infer the preferences. 
This assumption allows us to keep the bargaining game tractable, given our data limitation that we do not observe the exact nature of the interactions among the retiree and the firms.\footnote{We can replace the assumption with a weaker assumption that during the bargaining process, firms learn which are the two most competitive firms.} If $(\theta_i,\beta_i)$ were $i$'s private information, it would lead to a bargaining game with two-sided asymmetric information. We would then have to take a stand on the exact order of the moves, rules of information revelation, and updating, none of which would be informed by our data. 

We model the second round as an alternating offer bargaining process. The game's timing is as follows: In an arbitrary order, firms sequentially choose whether to improve their previous offer by a fixed amount $\varepsilon$ (play $Improve$) or to ``stay'' (play $Stay$). The process ends after a round in which all firms consecutively play $Stay$. Then the retiree chooses an offer. In Lemma \ref{lemma1}, we formalize the analysis with the proof in Appendix \ref{proofs}. 

Before we proceed, we introduce additional notations.
Let ${P}^{\max}_{ij}$ be the maximum $j$ can offer to $i$ without losing money, i.e., ${P}^{\max}_{ij}$ solves $C({P}^{\max}_{ij})={P}^{\max}_{ij} \times UNC_j = S_i$, or equivalently $1 = r_{ji} \times \bs_{i}^*({P}^{\max}_{ij})$. Let $j_i^*$ denote the firm that can offer the highest utility to $i$ without losing money, i.e., $
j_i^* := \arg\max_{j \in J} \rho_i({P}^{\max}_{ij})+\theta_i\times b_i({P}^{\max}_{ij})+{\beta_{i}\times{Z}_{j}}. 
$

\begin{lemma}\label{lemma1}
In the bargaining game, firm $j_i^*$ wins the annuity contract and, as $\varepsilon$ goes to zero, ends up paying a pension $\tilde P_{ij_i^*}$ such that 
\begin{eqnarray}
\beta_{i}\times{Z}_{j_i^*}+\theta_i b_i(\tilde P_{ij_i^*})+ \rho_i(\tilde P_{ij_i^*})= \max_{k\neq j^*_i}\Big\{\beta_{i}\times{Z}_{k} +\theta_i b_i({P}^{\max}_{ik})+ \rho_i({P}^{\max}_{ik})\Big\}.\label{eq:secondround}
\end{eqnarray}
Symmetric behavioral strategies that support this Perfect Bayesian Equilibrium are: 
\begin{enumerate}
\item For the retiree, choose whichever firm made the best offer (including non-pecuniary attributes), i.e., retiree $i$ chooses firm $j_i^*$ if 
$
j_i^* = \arg\max_{j \in J} \rho_i(\tilde P_{ij})+\theta_i\times b_i(\tilde P_{ij})+{\beta_{i}\times {Z}_{j}},
$
where $\tilde P_{ij}$ refers to the last offer of firm $j$ (or to its first-stage offer if it did not raise it during the bargaining game).
\item For a firm $j$, play $Improve$ iff $\tilde{\tilde P}_{ij}+\varepsilon <P_{ij}^{\max}$ and 
 $
\beta_{i}\times{Z}_{j}+\theta_i b_i(\tilde{\tilde P}_{ij})+ \rho_i(\tilde{\tilde P}_{ij}) < \max_{k\neq j}\Big\{\beta_{i}\times{Z}_{k} +\theta_i b_i(\tilde{\tilde P}_{ik})+ \rho_i(\tilde{\tilde P}_{ik})\Big\}, 
$
where $\tilde{\tilde P}_{ik}$ refers to the standing offer of firm $k$ (or to its first-stage offer when we are in the initial round of the bargaining game). \label{lemma:1}
\end{enumerate}
\end{lemma}

\section{Identification\label{section:identification}}
In this section, we study the identification of the conditional distribution of bequest preferences $F_{\theta|S}(\cdot|\cdot)$, the distribution of preferences for risk ratings $F_{\vs|S}(\cdot|\cdot)$, the conditional cost distribution $W_{r|S}(\cdot|S)$, and the channel and savings-specific information-processing cost $\alpha$. 

Our observations are the outcomes of buying annuities by $N$ retirees as outlined in the previous sections. 
In particular, for each retiree $i\in N$, we observe her socioeconomic characteristics $X_i = (\tilde{X}_i, S_i)$, set ${\bf A}_i$ of annuity products that she solicits offers for, the set of active firms (i.e., potential entrants) $\tilde{J}_i$ at the time of $i$'s retirement, the set of participating firms $J_i\subseteq \tilde{J}_i$, their risk ratings $\{Z_j: j=1,\ldots, J_i\}$ and their pension offers for each product and $i$'s final choice. Pursuant to our discussion in Section \ref{section:demand}, for each retiree and each annuity $a\in {\bf A}_i$ we use the offers and the Gompertz estimates of mortality to determine the expected present discounted utilities from the pensions ${\boldsymbol \bs}_{ia}:=(\bs_{1a},\ldots, \bs_{J_ia})$, and the expected present discounted utilities from the associated bequest ${\boldsymbol b}_{ia}:=(b_{1a},\ldots, b_{J_ia})$. Henceforth, when we refer to either pensions or bequests we mean these utilities.

\subsection{ Distribution of Bequest Preference \label{section:bequest}}
Here we study the identification of the conditional distribution of bequest preference, $F_{\theta|S}(\cdot|S)$, with support $[0, \overline{\theta}]$ by comparing chosen bequest and foregone bequests. For this purpose, it suffices to compare, for each retiree, the bequests offered by the winning firm in the first round, associated with different products. Focusing only on the winning firm allows us to bypass the need to know $\beta_{i}\times{Z}_{j_i^*}$. 
  
For intuition, let us suppose that there are only two types of annuities, $a\in {\bf A}=\{1,2\}$, such that (after relabeling, if necessary) annuity $a=1$ offers a smaller bequest (larger pension) than the annuity $a=2$. For notational ease, we can suppress the retiree and firm indices and hold the two fixed. Let $\chi\in\{1,2\}$ denote observed annuity choice and let $U_a$ denote utility from annuity $a\in {\bf A}$. 

A retiree chooses $a=1$, i.e., $\chi=1$ if and only if ${U}_{1}\geq U_{2}$, or equivalently $\theta \leq \frac{\rho_{1}-\rho_{2}}{b_{2}-b_{1}}=-\frac{\Delta \rho_{12}}{\Delta b_{12}}$, where we use $\Delta \rho_{aa'}$ and $\Delta b_{aa'}$ to denote the differences in pensions $(\rho_{a'}-\rho_{a})$ and bequests $(b_{a'}-b_{a})$, respectively. In other words, the bequest preference for a retiree who chooses low bequest is bounded above by the ``price" (i.e., pension) gradient. Then conditional probability that a retiree with characteristics $X$ chooses $a=1$ is 
\begin{eqnarray*}
\Pr(\chi=1|\tilde{X}, S)=F_{\theta|S}\left(-\frac{\Delta\rho_{12}}{\Delta b_{12}}\Big|S\right)=\zeta(S)+ (1-\zeta(S))\times \tilde{F}_{\theta|S}\left(-\frac{\Delta\rho_{12}}{\Delta b_{12}}\Big|S\right).
\end{eqnarray*} 
We estimate the left-hand side probability $\Pr(\chi=1|X)$ using logistic regression, and we also observe the pension gradient $\left(\frac{\Delta\rho_{12}}{\Delta b_{21}}\right)$, which is the argument of $F_{\theta|S}(\cdot|\cdot)$ on the right-hand side. 
Under our maintained exclusion restriction assumption that the non-saving characteristics $\tilde{X}$ affect choice probabilities through pensions and bequests but not the distribution $F_{\theta|S}(\cdot|\cdot)$, as the non-saving characteristics $\tilde{X}$ vary exogenously across retirees, the pension gradients vary allowing us to ``trace" the continuous part, $\tilde{F}_{\theta|S}(\cdot|\cdot)$ everywhere over $(0, \overline{\theta}]$.  
Formally, with rich variation in $\tilde{X}$, for a $t\in (0, \overline{\theta}]$ there exists a pair $\{ \Delta\rho_{12}, \Delta b_{21}\}$ such that $t =- \frac{\Delta\rho_{12}}{\Delta b_{12}}$, then the distribution $\tilde{F}_{\theta|S}(\cdot|\cdot)$ is nonparametrically identified. 

To identify the mass-point $\zeta$, we can use the subset where the price gradient approaches zero from the ``right side," i.e., $\frac{\Delta\rho_{12}}{\Delta b_{12}}\rightarrow 0^+$. The price gradient is close to zero when either because the two pensions $\rho_1$ and $\rho_2$ are close to each other or the difference in bequests $\Delta b_{12}$ is large, relative to the differences in pensions $\Delta \rho_{12}$. 
In other words, annuity $a=2$ has relatively cheaper bequest, and the only reason why a retiree would still choose annuity with low bequest is if she does not value bequest, i.e., $\theta\approx0$. Thus, we can identify the mass-point  
\begin{eqnarray*}
 \lim_{\frac{\Delta\rho_{12}}{\Delta b_{12}}\rightarrow 0^+} \Pr(\chi=1|\tilde{X}, S)=\lim_{\frac{\Delta\rho_{12}}{\Delta b_{12}}\rightarrow 0^+}\left( \zeta(S)+ (1-\zeta(S))\times \tilde{F}_{\theta|S}\left(-\frac{\Delta\rho_{12}}{\Delta b_{12}}\Big|S\right)\right)=\zeta(S), 
\end{eqnarray*}
where the last equality follows from $\lim_{\frac{\Delta\rho_{12}}{\Delta b_{12}}\rightarrow 0^+}\tilde{F}_{\theta|S}\left(-\frac{\Delta\rho_{12}}{\Delta b_{12}}\Big|S\right)=0$.

This identification strategy extends to cases with more than two types of annuities, i.e., $|{\bf A}|\geq2$. 
For the identification of $\tilde{F}_{\theta|S}(\cdot|\cdot)$ we can focus on the ``extreme" cases, such as the probability of choosing annuity with the smallest bequest and the probability of \emph{not choosing} the annuity with the largest bequest. 
We order the annuities offered by the winning firm (in either stages) in terms of the associated bequests, after relabeling if necessary, as, $b_1\leq b_2\leq\ldots\leq b_{|{\bf A}|-1}\leq b_{|{\bf A}|}$ and the corresponding pensions as $\rho_1\geq \rho_2\geq\ldots\geq \rho_{|{\bf A}|-1}\geq \rho_{|{\bf A}|}$. Let $\chi\in\{1,\ldots, |{\bf A}|\}$ be the annuity choice. Then the probability of choosing $b_1$ is 
\begin{eqnarray*}
\Pr(\chi=1|\tilde{X}, S)&=&\int_{\Theta}\mathbbm{1}\left\{U_1\geq U_{a}, a \in {\bf A} |\theta\right\}dF_{\theta|S}(\theta|S)=F_{\theta|S}\left( \min_{a\in {\bf A}} \left\{-\frac{\Delta \rho_{1a}}{\Delta b_{1a}}\right\}\Big|S\right),
\end{eqnarray*}
and the probability of \emph{not choosing} the annuity with the \emph{largest} bequest, i.e.,$\chi\neq |{\bf A}|$, is  
\begin{eqnarray*}
	\Pr(\chi\neq |{\bf A}| \Big|\tilde{X}, S)=F_{\theta|S}\left( \max_{a\in{\bf A}} \left\{-\frac{\Delta \rho_{|A|a}}{\Delta b_{|A|a}}\right\}\Big|S\right). 
\end{eqnarray*}
Then we can use $\{\Pr(\chi=1|\tilde{X}, S), \Pr(\chi\neq |{\bf A}||\tilde{X}, S)\}$ to identify $\tilde{F}_{\theta|S}(\cdot|\cdot)$, where, as mentioned above, the identifying sources of variation are, $\tilde{X}$, annuitization costs across firms, and the number of participating firms, which in turn lead to variations in pensions and bequests. To identify the mass-point, we can rely on the same limit-argument as above. 

\subsection{Information-Processing Cost}
Let ${\mathcal J}$ denote the unique values of $J_i$ across all $i\in N$. 
Consider the subset of retirees with $|J_i|=J$.
Then we can identify the choice probability for $j=1,\ldots, J
+1$, given ${X}={x}$, $Z=z$ and $({\boldsymbol \bs}, {\bf b})$, by  
{\footnotesize\begin{eqnarray}
\hat{\sigma}_{j}({x},z,{\boldsymbol \bs}, {\bf b}|J)= \sum_{i}\frac{\mathbbm{1}[D_{i}^{1}=j, {X}_{i}={x}, Z=z, {\boldsymbol \bs}, {\bf b}]}{\sum_{i}\mathbbm{1}[{X}_{i}={x}, Z=z, {\boldsymbol \bs}, {\bf b}]}; \quad
\hat{\sigma}_{J+1}( \tilde{x},z,{\boldsymbol \bs}, {\bf b}|J)=1-\sum_{j=1}^{J}\hat{\sigma}_{j}({x},z,{\boldsymbol \bs}, {\bf b}),\quad \label{eq:prob-hat}
\end{eqnarray}}
\noindent where $D_i^1=j$ denotes $i$ choosing firm $j$. 
Applying (\ref{eq:prob-hat}) to the relevant subsample, we can identify $\{{\sigma}_{j}({x},z,{\boldsymbol \bs}, {\bf b}|J)\}_{j\in J}$ for all $J\in {\mathcal J}$. 
We can also identify the probability that there are $J$ participating firms as $p(J)=\#\{\texttt{retirees with $J_i=J$}\}/N$, and together we identify ${\sigma}_{j}({x},z,{\boldsymbol \bs}, {\bf b})=\sum_{J\in{\mathcal J}} {\sigma}_{j}({x},z,{\boldsymbol \bs}, {\bf b}|J)\times p(J)$. 
Integrating (\ref{eq:prob}) with respect to $F_{\vs}$ and using the definition of $\hat{\sigma}_{j}({x},z,{\boldsymbol \bs}, {\bf b})$ gives
\begin{eqnarray}
\hat{\sigma}_{j}(x,z,{\boldsymbol \bs}, {\bf b})=\int\frac{\exp\left(\log \sigma_{j}^{0}+\frac{U_{ij}}{\alpha}\right)}{\sum_{k=1}^{J}\exp\left(\log \sigma_{k}^{0}+\frac{U_{ij} }{\alpha}\right)+\exp\left(\frac{\E U_{i}}{\alpha}\right)}dF_{\vs}(\vs).\quad\label{eq:mixture}
\end{eqnarray}
Taking the derivative of (\ref{eq:mixture}) with respect to $\bs_{j}$ identifies the cost 
$\alpha = \frac{{\sigma}_{j}({x},z, {\boldsymbol \bs}, {\bf b}) (1- \hat{\sigma}_{j}({x},z,{\boldsymbol \bs}, {\bf b}))}{{\frac{\partial \hat{\sigma}_{j}({x},z,{\boldsymbol \bs}, {\bf b})}{\partial \bs_{j}}}}$.\footnote{To estimate $\alpha$, we use a logit specification to model the LHS of (\ref{eq:mixture}) so the derivatives are well defined.}

Thus, the information-processing cost depends on the choice probability elasticity with respect to $\rho$. 
For intuition, consider an extreme case when the choice for $j$ is insensitive to changes in premium, i.e., ${\frac{\partial \hat{\sigma}_{j}({x},z,{\boldsymbol \bs})}{\partial \bs_{j}}}\approx0$, then it implies that $\alpha \approx +\infty$ because the only way to rationalize the fact that retirees do not respond to changes in pension is that their information-processing cost is extremely large. 
If the demand is elastic with respect to the pensions, then the cost of processing information is low, and vice versa. 
We can apply this strategy to the appropriate subsample to allow this cost to vary with channel and savings. 
 
\subsection{ Risk-Rating Preferences and Annuitization Costs }
For the identification of the preference distribution $F_{\vs|S}$ and the cost distribution $W_{r|S}$, it is sufficient to consider only those who buy annuities in the second round, where the chosen pension and bequests satisfy (\ref{eq:secondround}).
Let $\tilde P_{ij_i^*}$ be the chosen offer, then from (\ref{eq:secondround}) we get 
\begin{eqnarray}
\rho_i(\tilde P_{ij_i^*})+\theta_i b_i(\tilde P_{ij_i^*})= \max_{k\neq j^*_i}\Big\{\beta_{i}\times{Z}_{k} +\theta_i b_i({P}^{\max}_{ik})+ \rho_i({P}^{\max}_{ik})\Big\}-\beta_{i}\times{Z}_{j_i^*}.\label{eq:secondroundid}
\end{eqnarray}
Let $k_i^*$ denote the runner-up firm for retiree $i$. We do not directly observe $k_i^*$, but we can use first-round offers to identify the runner-up firm for the identification. In particular, if the runner-up in the first round is one of the two most competitive firms in the second round, then we can determine $k_i^*$. The runner-up firm in the first round is the firm with the highest probability of being chosen in the first round once we exclude the winning firm $j_{i}^*$.
If we can estimate the probability of being chosen in the first round, we can identify the runner-up firm for the second round. We detail the probability estimation steps in Appendix \ref{section:runnerup}, and for the rest of the section, we treat $k_i^*$ as known.

Using the definition of $k_i^*$, we can simplify the right-hand side of (\ref{eq:secondroundid}) and re-write it as  
\begin{eqnarray}
\rho_i(\tilde P_{ij_i^*})+\theta_i b_i(\tilde P_{ij_i^*})&=& \beta_i\times({Z}_{k_i^*}-{Z}_{j_i^*})+ \theta_i b_i({P}^{\max}_{ik_i^*}) + \rho_i({P}^{\max}_{ik_i^*})\notag\\
&=& \beta_i\times({Z}_{k_i^*}-{Z}_{j_i^*})+ \underbrace{\mathfrak{U}({P}^{\max}_{ik_i^*},B_{ik_i^*}({P}^{\max}_{ik_i^*}); \theta_i)}_{\equiv\varpi_i}\notag\\&\equiv&\beta_i\times({Z}_{k_i^*}-{Z}_{j_i^*})+\varpi_i,
\qquad\label{eq:OLS}
\end{eqnarray}
where for the third equality we substituted the highest gross utility that the runner up firm $k_i^*$ can offer to retiree $i$ with $\varpi_i$. The variable $\varpi_i$ depends on $k_i^*$'s cost, which is unobserved, we can treat $\varpi$ as a random variable with a distribution $F_{\varpi}$. Our objective is to first identify $F_{\beta|S}$ and $F_{\varpi}$, and then identify the cost distribution $W_{r|S}$. 

\begin{figure}[t!]
\caption{\bf Histogram of Winner and Runner Up}
\centering
\includegraphics[scale=0.4]{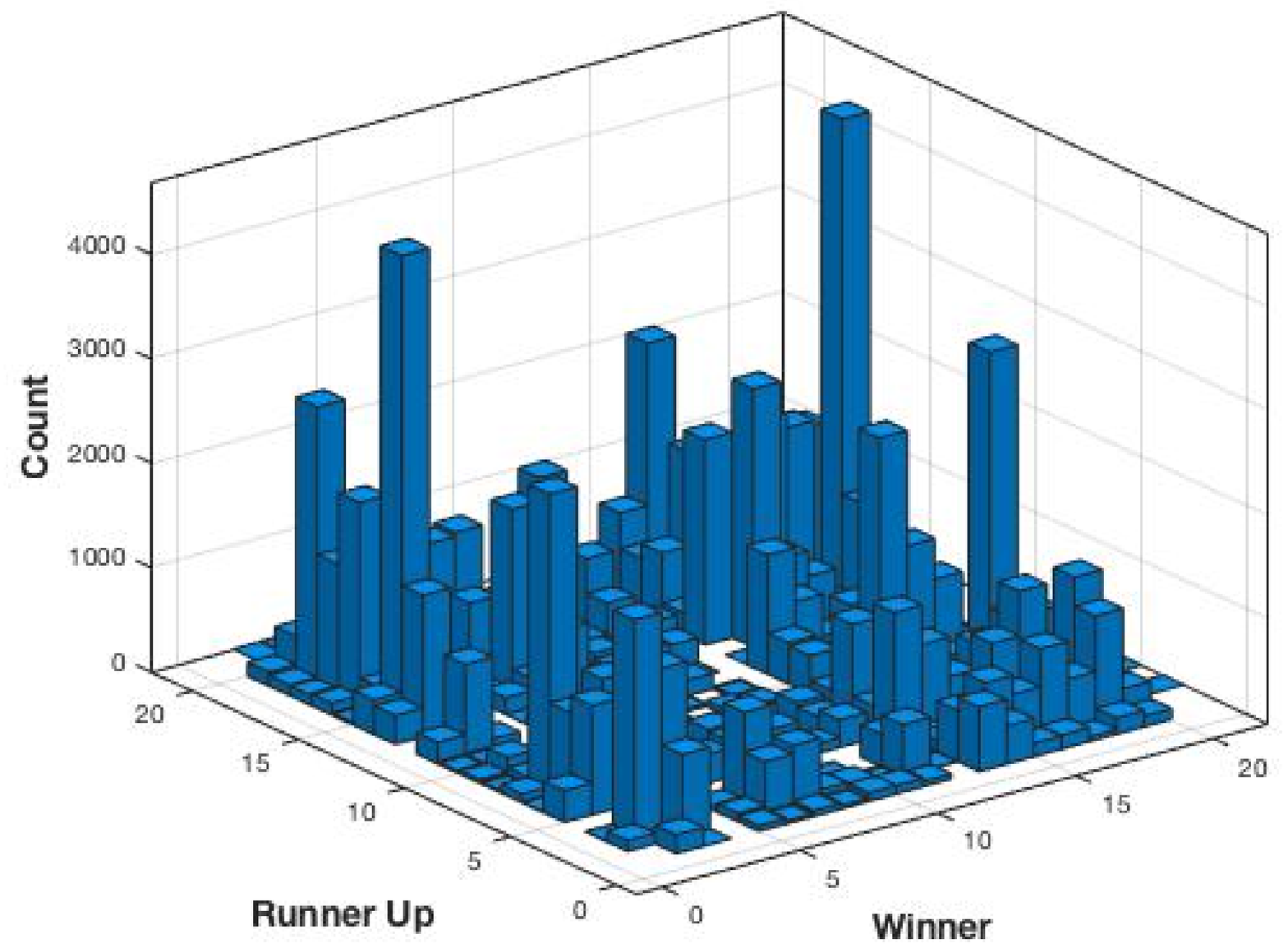}
\label{fig:Histogram}
\caption*{\footnotesize {\bf Note:} The histogram shows the identity of the winning firms (on the x-axis) and the runner-up firms (on the y-axis). The runner-up firm for a retiree is the firm that has the highest probability of being chosen by the retiree after excluding the chosen firm. The probabilities are estimated in Appendix \ref{section:runnerup}.}
\end{figure}

Suppose we observe individual-level $\theta_i$.
Then using (\ref{eq:OLS}) we can identify $F_{\beta|S}(\cdot|\cdot)$ as follows. First, using the chosen annuity we can determine the left-hand side of (\ref{eq:OLS}). Then, we can treat $\varpi$ as an ``error" in the random coefficient model (\ref{eq:OLS}). Third, we note that $({Z}_{k_i^*}-{Z}_{j_i^*}), \beta_i$ and $\varpi_i$ are uncorrelated. Moreover, firms' risk ratings do not change, but the retiree-specific difference $({Z}_{k_i^*}-{Z}_{j_i^*})$ vary across retirees because the identity of the two most competitive firms $(j_i^*, k_i^*)$ vary across retirees; see Figure \ref{fig:Histogram}. Then the identification follows from the application of the results in \cite{HoderleinKlemelaMammen2010}.  

However, we only know $F_{\theta|S}(\cdot|\cdot)$ and not the retiree-specific $\theta_i$. So we need additional assumptions on $F_{\beta|S}(\cdot|\cdot)$.
We assume that $F_{\beta|S}(\cdot|\cdot)$ is a normal distribution. 
To allow the parameters of $F_{\beta|S}(\cdot)$ to depend on observed characteristics in a flexible manner, we divide retirees into several groups, and estimate group-specific mean and variance of $\beta$. In particular, we create groups, which we denote by $G$, based on gender (male or female), three age groups (those who retire, before, at, or after the normal retirement age), five savings quintiles, and three channels (AFP, sales agents, or independent advisor), which gives us a total of 90 groups. 
Let $g(i)\in G$ denote $i$'s group, and let ${\beta}_i\sim {\mathcal N}(\beta_{g(i)}, \sigma_{g(i)}^2)$, where $(\beta_{g}, \sigma_{g}^2)$ are group $g$-specific mean and variance. 
We also assume that the savings affect the annuitization costs through the savings quintiles $S_{q}$, i.e., $r\sim W_{r|S}(\cdot|S_q)$, where $S_q$ is the $q^{th}$-quintile. 

Let $N_{q,J}$ denote the subset of retirees in the $q^{th}-$quintile that have $J\in\{13,14,15\}$ potential bidders.  
In our first step, we generate $L\times |N_g(i)|$ matrix that consist of i.i.d. draws of $\theta$ from ${F}_{\theta}(\cdot|S_q)$, where $L$ is a large number, and for our purpose we set $L=10,000$, and $N_g(i)$ is the set of retirees who belong to the same group as $i$. 
Second, for each $\ell^{th}$ row of this simulated matrix, which we denote by $\{\theta_i^\ell, i=1,\ldots, N_{g(i)}\}$, we estimate $\{\hat{\beta}_{g}^{\ell}: g=1,\ldots, G\}$ applying the method from \cite{Swamy1970} to 
 \begin{eqnarray}
{\bs}_{j_i^*}+\theta_i^{\ell}\times b_{j_i^*}= {\beta}^{\ell}_{g(i)}\times ({Z}_{k_i^*}-{Z}_{j_i^*})+ \varpi_{k_i^*}^{\ell}. \label{rcequation1}
\end{eqnarray}
Estimating (\ref{rcequation1}) $L=10,000$ times gives us group-specific estimate $\{\hat{\beta}_{g}^{\ell}: \ell=1,\ldots,L, g\in G\}$. Third, we can average across these estimates to get $\hat{\beta}_{g}=L^{-1}\sum_{\ell=1}^L \hat{\beta}_{g}^{\ell}$.\footnote{Although we did not formally determine the consistency of our approach, we performed several Monte Carlo experiments and found that for $L=10,000$ the average estimator performed well.} 

Next, we can show that conditional on $\{F_{\beta|S}, F_{\theta|S}\}$ we can identify $W_{r|S}(\cdot|\cdot)$.  
To this end, first note that the utility from the winning firm, which is given by the left-hand side of (\ref{eq:OLS}), is the second-largest (maximum) utility from $J$ firms. We observe the chosen pension and bequest, and so we know the distribution of the left-hand side of (\ref{eq:OLS}). Second, using the distribution of the left-hand side of (\ref{eq:OLS}) we can identify the parent distribution of the right-hand side of (\ref{eq:OLS}). Third, then we can use the fact that this distribution we recovered in the second step is a convolution distribution of $(\beta_i\times({Z}_{k_i^*}-{Z}_{j_i^*}))$ and $\varpi_i$. We can then identify the distribution of $\varpi_i$ using a deconvolution method, which is widely used in the classic measurement error literature (e.g., \cite{Schennach2016}). Finally, we can exploit the one-to-one mapping from $\varpi_i$ to $P_{ik_i^*}^{\max}$ --the maximum pension runner-up firm $k_{i}^*$ can offer to retiree $i$ as shown in Equation (\ref{eq:Pension}). This relationship with the definition of maximum (or break-even) pension, i.e., $C({P}_{ik_i^*}^{\max})=S_i$, identifies the distribution of $r=\frac{UNC_{k_i^*}}{UNC_i} = \frac{S_i}{{P}_{ik_i^*}^{\max}\times UNC_i}$. We formalize these four steps below, with the proof in Appendix \ref{proofs}. 

\begin{lemma}
$W_{r|S}(\cdot|\cdot)$ can be nonparametrically identified from $\{F_{\beta|S}(\cdot|\cdot), F_{\theta|S}(\cdot|\cdot)\}$.\label{lemma:Wr}
\end{lemma}

{\bf Selective Entry.} Let $\tilde{J}$ be the set of companies interested in selling annuities to $i$ with characteristics $X_i$. When $i$ requests an offer for a product, company $j\in\tilde{J}$ observes its cost $r_j$, and all firms simultaneously, decide whether to participate, and the cost (the same) $\kappa_i\geq0$ for each company to participate. This cost captures the opportunity cost to participate, and it can vary across retirees. Let $J\subset \tilde{J}$ denote the set of participating companies. All the firms that participate simultaneously make their offers. 

Under the symmetric Perfect Bayesian-Nash equilibrium, the entry decision is characterized by a unique threshold $r^*\in[\underline{r}, \overline{r}]$ such that firms participate only if their costs are less than $r^*$. Then the cost distribution among the participating firms is $W_{r|S}^*(r|S; \tilde{J}):=W_r(r|r\leq r^*, S; \tilde{J})=\frac{W_{r|S}(r|S)}{W_{r|S}(r^*|S;\tilde{J})}$.  
Let $r_{\tilde{J}}^*$ be the threshold with $\tilde{J}$ potential bidders, and suppose $\tilde{J}\in{\mathcal J}:=\{\underline{J},\ldots, \overline{J}\}$, where $\overline{J}$ is the maximum number of potential bidders and $\underline{J}$ is the smallest number of potential bidders. All else equal, $r_{\tilde{J}}^*$ decreases with $\tilde{J}$, so $W_{r|S}(r|S)$ is identified on the support $[\underline{r}, r_{\underline{J}}^*]$. 

In practice, to estimate $W^{*}_{r|S}(\cdot|S_q)$, we focus on the sub-sample of retirees with the top two firms with the same risk ratings. 
In our sample, close to 60,000 retirees are in this group and have $({Z}_{k_i^*}-{Z}_{j_i^*})=0$.
Substituting this in (\ref{rcequation1}) for ${J}\in\{13,14,15\}$ gives 
 \begin{eqnarray}
{\bs}_{j_i^*}+\theta_i^{\ell} b_{j_i^*}= \varpi_{k_i^*}^{\ell}, 
\label{rcequation2}
\end{eqnarray}
where the left-hand side is the known winning utility and the right-hand side is the unobserved maximum utility the runner-up firm can offer without incurring loss. 
From the estimated distribution of $({\bs}_{j_i^*}+\theta_i^\ell b_{j_i^*})$, we can estimate the parent distribution of $\hat{r}_{ji}$, i.e., $W_{r|S}^*(\cdot|S_q, {J})$ using kernel density estimation.

\section{Estimation Results\label{section:estimation}}
\subsection{\bf Preferences for Bequests.} 
\begin{figure}[ht!]
	\caption{\bf Estimated Distributions of Bequest Preferences}
	\centering
	\includegraphics[scale=0.32]{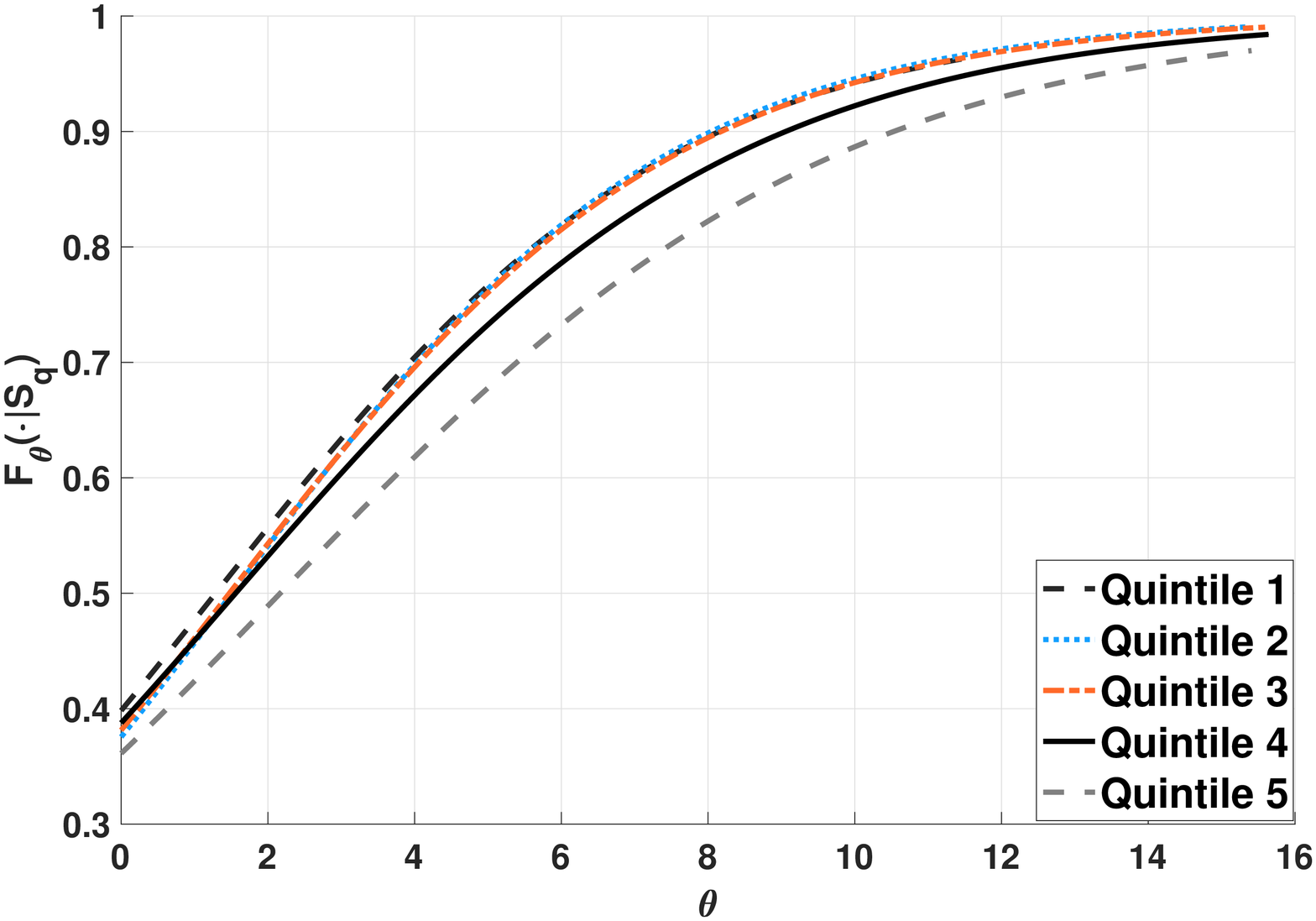}
	\label{fig:theta_by_quintile}
	\caption*{\footnotesize {\bf Note:} This figure displays estimated conditional distribution of preference for bequests $F_{\theta}(\cdot|S_q)$ given savings quintile $S_q, q=1,\ldots, 5$, as we move from the left to the right. }
\end{figure}

In Figure \ref{fig:theta_by_quintile}, we display the estimates of the conditional distributions of preferences for bequests, given savings quintiles $\{\hat{F}_{\theta|S}(\cdot|S_q), q=1,\ldots, 5\}$. 
Our estimates suggest that approximately 40\% of retirees do not value leaving bequests generating a positive mass at $\theta=0$. 
However, we find that the mass points do not vary across savings quintiles, although the mass point for the highest savings quintile is the smallest one.
In Table \ref{table:thetamean}, we present some summary statistics associated with these conditional distributions. 
The variance and the inter-quartile range shown in the last two columns of Table \ref{table:thetamean} suggest considerable heterogeneity within and across the savings quintiles.

That the conditional distributions in Figure \ref{fig:theta_by_quintile} ``shift right" with saving quintile suggests that the average bequest preference increases with savings. 
As a result, the mean $\E(\theta|S_q)$ in Table \ref{table:thetamean} increases with savings quintiles, which is consistent with the hypothesis that with decreasing marginal utility from a pension, the marginal utility of bequest for an altruist retiree increases with savings.

\begin{table}[ht!]
	\caption{\bf Summary Statistics of Preference for Bequests \label{table:thetamean}}
\begin{center}
\begin{tabular}{lllll}
\toprule
{\bf Savings} & {\bf Mean} & {\bf Median} & {\bf Std. Dev.}& {\bf IQR} \\
\toprule
Q1 & 2.784 & 1.274 & 3.526 & 4.690 \\
Q2 & 2.899 & 1.505 & 3.597 & 4.773 \\
Q3 & 2.971 & 1.588 & 3.681 & 4.901 \\
Q4 & 3.208 & 1.575 & 4.053 & 5.290 \\
Q5 & 3.812 & 2.166 & 4.567 &6.317 \\  
\bottomrule
\end{tabular}
	\caption*{\footnotesize {\bf Note:} Mean, median, standard deviation and inter-quartile range of preference for bequests, by saving quintiles. These statistics are calculated using simulated $\theta$ from $\{\hat{F}_{\theta|S}(\cdot|S_q)\}_{q=1}^5$ as shown in Figure \ref{fig:theta_by_quintile}.}
\end{center}
\end{table}

\subsection{\bf Risk Ratings and Information-Processing Costs.}
\begin{figure}[h!]
	\caption{\bf Group-Specific Mean Risk Ratings Preference\label{fig:beta}}
	\begin{center}
	\includegraphics[scale=0.5]{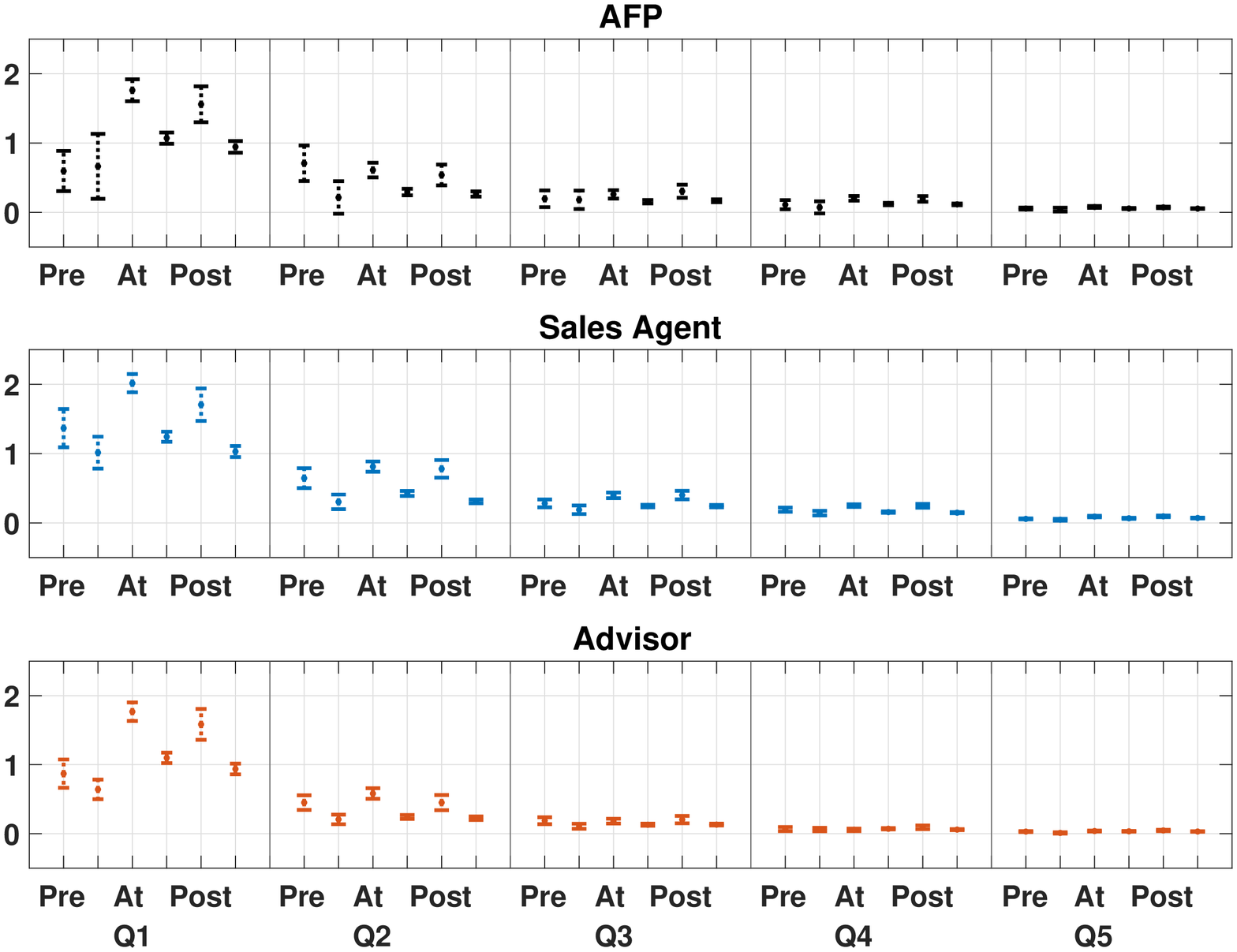}
	\label{fig: betas}
	\caption*{\footnotesize {\bf Note:} These figures display the estimates for group-specific mean of $\E(\beta_g)$, from (\ref{eq:OLS}). Each panel (row) corresponds to a channel, and each channel is divided into five quintiles. And within each channel-quintile box, parameters are ordered by retirement age (before, after or at NRA), and for each age group, the two estimates correspond to male and a female respectively. The two bars represent 95\% confidence intervals.}
	\end{center}
\end{figure}

Next, we present the estimates of the preference for risk rating. Figure \ref{fig:beta} displays the mean $\hat{\beta}_{g}$, for all 90 groups, with their corresponding 95\% bootstrapped confidence intervals.
These estimates suggest that those in the lowest two savings quintiles are the only ones who significantly value firms' risk ratings. Furthermore, among those who value risk ratings positively, males exhibit slightly stronger preference than females, although the differences across gender are statistically insignificant in many cases. 

On the face of it, that preference for risk ratings decreases with savings might seem counter-intuitive. There are at least two reasons why it is so. First, if the risk rating is a proxy for financial health, there should be no heterogeneity across retirees. Second, because of the government's guarantee, those with higher savings are more exposed to the ``bankruptcy risk" than those with lower savings, so the high savers should care more about the risk ratings than the low savers. 
One way to rationalize this finding is to consider the information-processing cost. If these groups have different information-processing costs, then the rational-inattention model suggests that the group with lower information processing learns more. 

In Table \ref{tab:lambda-median}, we present the estimated group-specific information-processing costs ($\hat{\alpha}_g$).
We find that the cost decreases with savings, and the absolute decrease is most prominent among the retirees with the lowest quintile and who have sales agents because those with higher savings tend to be more educated. 
So, even if the prior suggests that risk ratings are important, those with lower information-processing costs revise their preferences downwards to reflect that bankruptcy is a rare event in Chile and that most firms have good ratings.

\begin{table}[h!]\centering \caption{\bf Information-Processing Cost \label{tab:lambda-median}}
\scalebox{0.85}{\begin{tabular}{lllll}
\toprule
{\bf Savings} & {\bf AFP} & {\bf Sales Agent} & {\bf Advisor} & {\bf Full Sample} \\
\toprule
Q1   &0.009		& 0.027		&0.006	&0.021 \\
Q2   &0.006		&0.019		&0.004	&0.016 \\
Q3   &0.005		&0.013		&0.003	&0.013 \\
Q4   &0.005		&0.012		&0.003	&0.005 \\
Q5   &0.005		&0.012		&0.003	&0.006 \\
\midrule
{\bf Overall}  &0.005		&0.013		&0.003	&0.009 \\
\bottomrule
\end{tabular}}
	\caption*{\footnotesize {\bf Note:} Estimates of the median of information-processing cost, by savings quintiles and intermediary channel. }
\label{resultslambda}
\end{table}

\subsection{\bf Annuitization Costs.}
\begin{figure}[ht!]
\caption{\bf Conditional Distributions of Annuitization Costs \label{fig:Ww}}
\begin{subfigure}{.5\textwidth}
 \centering
 \includegraphics[scale=0.25]{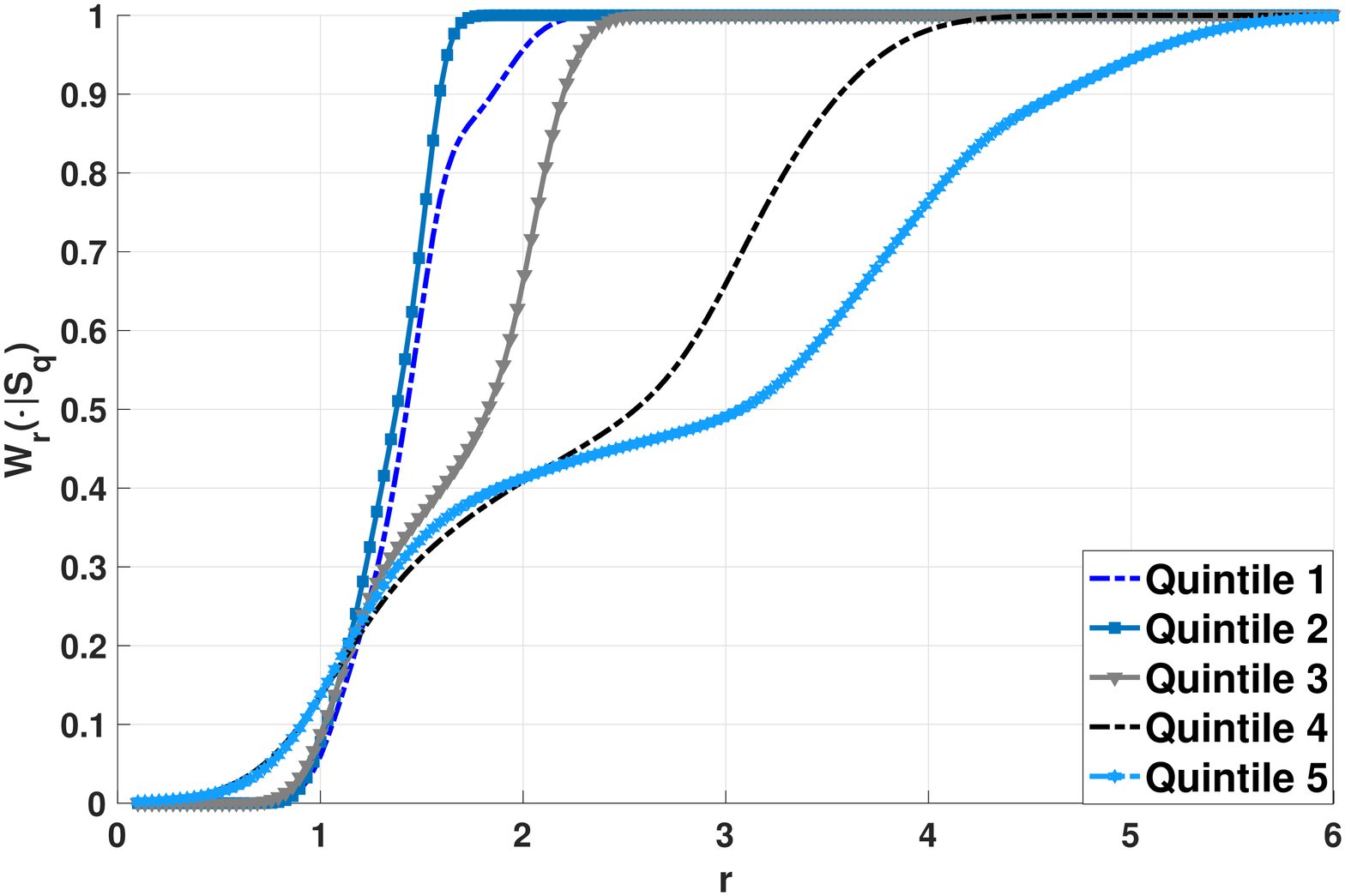}  
 \caption{Over the Full Support}
 \label{fig:W}
\end{subfigure}
\begin{subfigure}{.5\textwidth}
 \centering
 \includegraphics[scale=0.25]{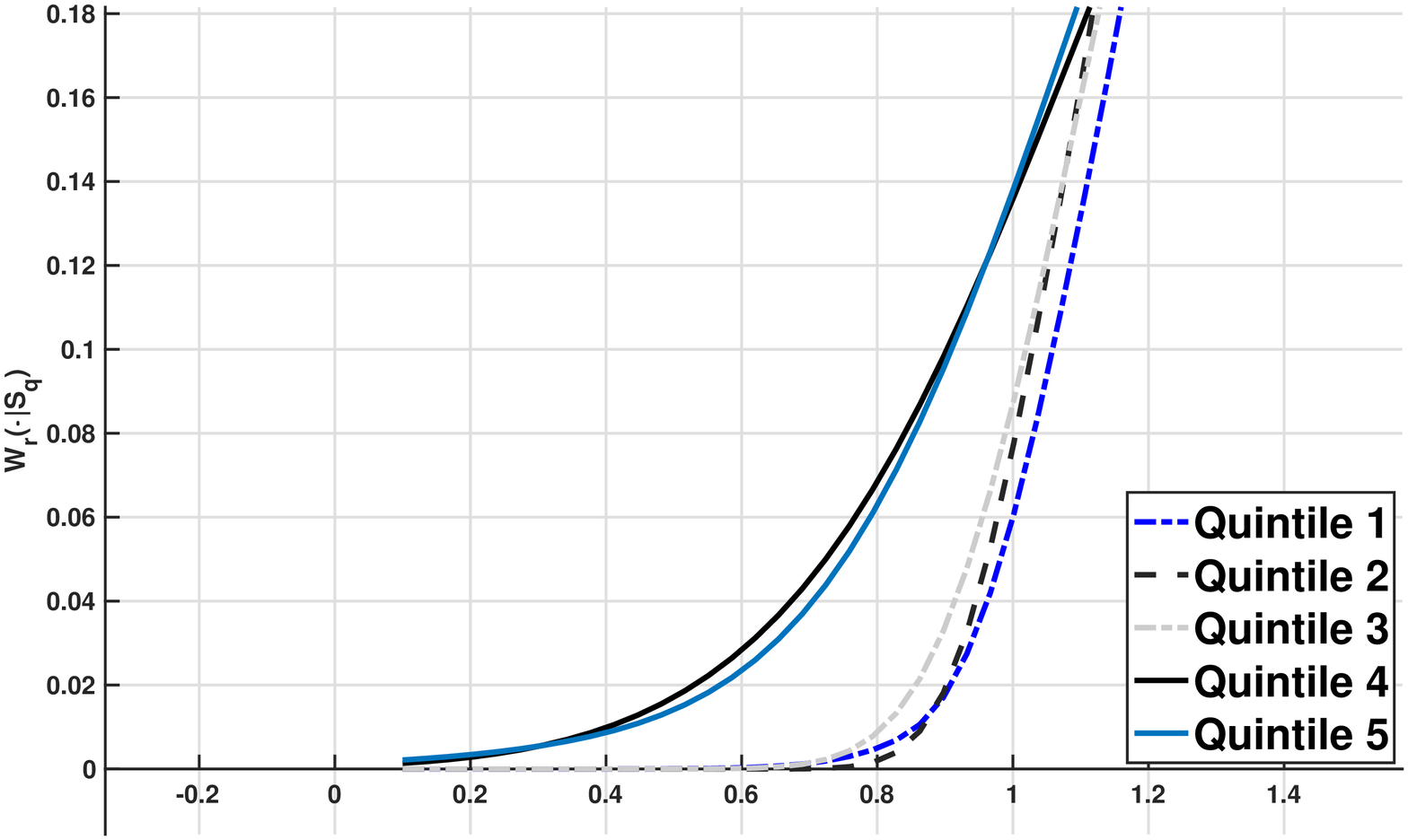}  
 \caption{ Focusing on $r<1$.}
 \label{fig:W_}
\end{subfigure}
\caption*{\footnotesize {\bf Note:} The first sub-figure shows the estimated conditional distribution of relative annuitization costs, by savings quintiles. The second figure shows focuses only on the support where $r<1$. }
\end{figure}

Figure \ref{fig:Ww} presents the estimates of the conditional distributions of costs given the savings quintile. 
 Recall that $r_{ji}=\frac{UNC_j}{UNC_i}$ is the ratio of firm $j$'s UNC to retiree $i$'s UNC and is thus unit free.\footnote{ Recall that we work with $r_{ji}$ instead of $UNC_j$ because each retiree is unique and has different mortality and normalizing $UNC_j$ by $UNC_i$ homogenizes cost across retirees. Then we can ``pool" the data from different retiree auctions together and make inter-retiree comparisons.}
 So, $r_{ji}>1$ means that firm $j$'s cost of annuitizing $i$'s savings is above the average market return rate.  
  
As we can see from Figure \ref{fig:W}, because the distributions ``shift right" with savings, it suggests that the annuitization cost also increases with savings. 
In Table \ref{table:r}, we present the summary statistics of $r$ by savings quintiles, and we can see that the mean annuitization costs do not seem to increase with savings quintiles.  

The shapes of the distributions when $r<1$ (see Figure \ref{fig:W_}) can explain this pattern.
Firms are twice as likely (14\% versus 6\%) to have $r<1$ when the retiree belongs to the top two savings quintiles than when they do not. Thus, this ``crossing" of the conditional distribution functions lowers the overall average costs for high savers. 

\begin{table}[ht!]
	\caption{\bf Summary Statistics of $r$\label{table:r}}
\centering
\scalebox{0.85}{\begin{tabular}{lllll}
\toprule
{\bf Savings} & {\bf Mean} & {\bf Median} & {\bf Std. Dev.}& {\bf IQR} \\
\midrule
Q1 & 2.74 & 3.1 & 1.47 & 2.7 \\
Q2 & 2.75 & 3.11 & 1.47 & 2.7 \\
Q3 & 2.73 & 3.07 & 1.46 & 2.69 \\
Q4 & 2.77 & 3.12 & 1.47 & 2.69 \\
Q5 & 2.76 & 3.12 &1.48&2.72 \\  
\bottomrule
\end{tabular}}
	\caption*{\footnotesize{\bf Note:} The table displays mean, median, standard deviation and inter-quartile range of the annuitization costs $r$. These statistics are calculated using simulated $r$ from $\{\hat{W}_{r|S}(\cdot|S_q)\}_{ q=1}^5$ as shown in Figure \ref{fig:W}.}

\end{table}

In equilibrium, the lowest two order-statistics of the cost determine the pensions, which in turn depends on the left tail of the distributions (Figure \ref{fig:W_}).
So, firms are twice as likely to have $r<1$ for the highest two quintiles than the other three quintiles, translating into a more significant gap between what firms offer and their break-even offer. The best way to illustrate this is to use the estimated cost distributions and determine the maximum pension that firms can offer without losing money, i.e., the break-even pension.  

\begin{figure}[ht!]
	\caption{\bf Distributions of Maximum Pension $P^{\max}$} 
	\centering
	\includegraphics[scale=0.4]{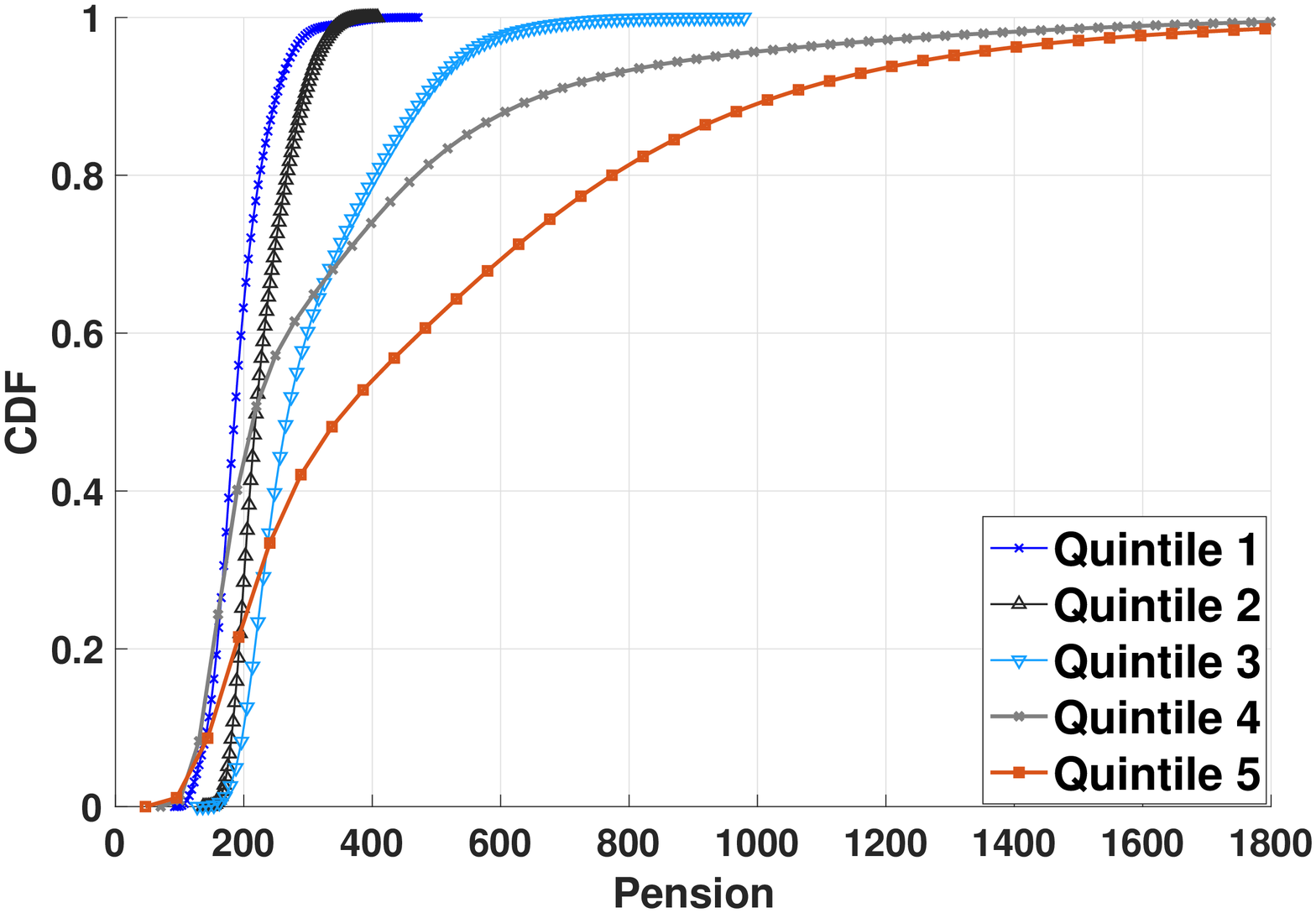}
	\label{fig:maximumP}
	\caption*{\footnotesize {\bf Note:} Conditional distributions of maximum pensions for retiree with median savings within each quintile. For each savings quintile $1\leq q\leq 5$, we simulate several $r$s from $\hat{W}_{r|S}(\cdot|S_q)$ displayed in Figure \ref{fig:W}, and we determine the median savings among this group. Using these $r$s and the median saving, we determine the maximum pensions $P^{\max}$ that firms can offer without making loss and estimate the distribution of $P^{\max}$. }
\end{figure}

To this end, we run the following simulation exercise: (i) for each savings quintile, we identify the retiree with the median income (among this subsample); (ii) simulate $\{r^{(\ell)}: \ell=1, \ldots, 1000\}$'s from the relevant distribution $\hat{W}_{r|S}(\cdot|\cdot)$; (ii) using the savings and the estimated $UNC_i$ of the retiree identified in step (i), for each draw $r^{(\ell)}$ determine $UNC_j$ and from that the maximum pension is given by the zero-profit condition, i.e., $P_{ij} = \frac{S_i}{UNC_j}$. 
Figure \ref{fig:maximumP} shows the resulting distributions of these pensions.
We can see, those firms would offer more for retirees with higher savings than for those with lower savings (per dollar). 

\section{Counterfactual Results \label{section:counterfactual}}
In this section, we evaluate the effect of asymmetric information on equilibrium pensions and retiree's welfare. Then we consider ways to improve the market by simplifying the current system, replacing it with a standard English auction, removing risk ratings from the supply side to increase competition (by selecting the firm that pays the highest pension), and automating the system, so retirees do not use risk ratings to choose a firm. 
We present pensions and gross utility under the current system, under complete information, and English auctions. 

\subsection{Complete Information}
We begin by considering the effect of asymmetric information on pensions and welfare and how they vary across savings quintiles, and the intensity of competition (i.e., the number of potential bidders). 
To determine pensions under complete information, we divide retirees into 15 groups based on their savings quintiles and their corresponding number of potential bidders. Then, for each retiree in a group, we use the appropriate $\hat{W}_{r|S}(\cdot|S_q)$ to determine the cost $r$ for the potential firms and determine the lowest cost among those draws. The winner will be the bidder with the lowest cost. Then, we determine the break-even pension the winner can offer. We repeat this step 10,000 times for each retiree and determine the average pension. 
 
\begin{figure}[ht!]\caption{\bf Estimated Distributions of Pensions (in 1000s of US\$) \label{fig:cf_cdf_pensions}}
 \centering
 \includegraphics[scale=0.4]{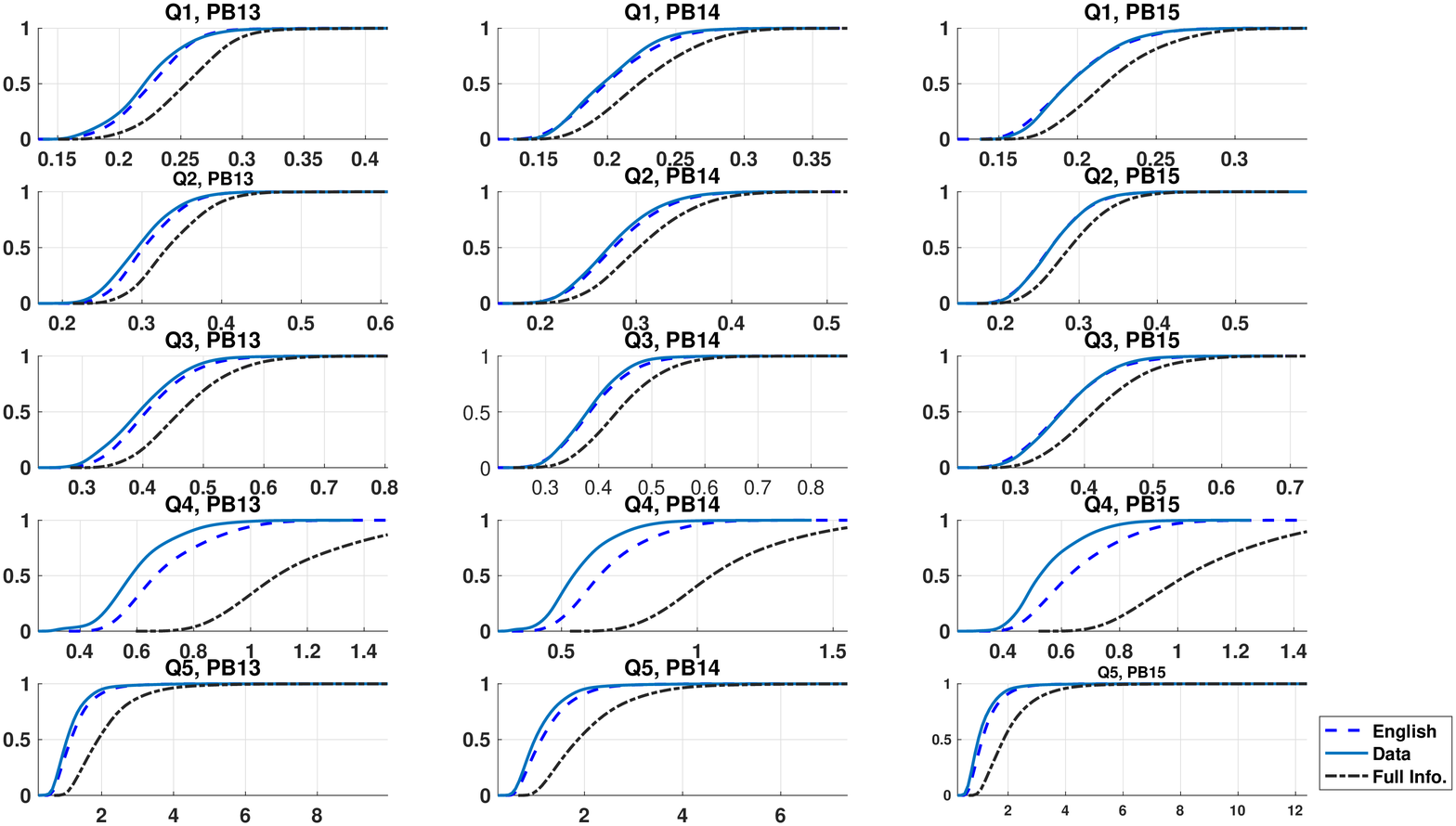}  
\caption*{\footnotesize {\bf Note:} Distributions of pensions (in thousands of U.S. dollars) under the current system (solid blue), under English auction (dashed blue) and under full information (dotted red), by savings quintiles (rows) and the number of potential bidders (columns). The sample includes only those who choose in the second round.}
\end{figure}
 
In Figure \ref{fig:cf_cdf_pensions}, we present the distributions of chosen pensions and the pension under the counterfactual of complete cost information. 
As expected, the pension distribution under complete information, first-order stochastically dominates the observed pensions' distribution.
Interestingly, the gap between the two distributions is substantial for those with higher savings. 

In Table \ref{table:pension_full_info}, we present the mean and median of current pensions as a percentage of the complete information pensions for each group. We find that for the lowest three savings quintiles, the numbers are at least 85\%, whereas for the top-two saving quintiles, they are significantly lower. 
These results are consistent with the shape of the cost distributions in Figure \ref{fig:cf_cdf_pensions}. 
 
 \begin{table}[ht!]
\begin{center}
\caption{\bf Pensions under Current and English Auctions, Relative to Full Info. \label{table:pension_full_info}}
\scalebox{0.75}{\begin{tabular}{llll}
\toprule
{\bf Savings} \textbackslash {\bf Potential Bidders} & 13    & 14    & 15    \\
\midrule
Q1   & \begin{tabular}[c]{@{}l@{}}(87\%, 87\%) \\ (88\%, 88\% )\end{tabular} & \begin{tabular}[c]{@{}l@{}}(91\%, 91\%)\\ (89\%, 89\%)\end{tabular} & \begin{tabular}[c]{@{}l@{}}(93\%, 93\%) \\ (89\%, 89\%)\end{tabular} \\
\hline
Q2   & \begin{tabular}[c]{@{}l@{}}(89\%, 90\%)\\ (90\%, 90\%)\end{tabular} & \begin{tabular}[c]{@{}l@{}}(94\%, 93\%)\\ (91\%, 91\%)\end{tabular} & \begin{tabular}[c]{@{}l@{}}(97\%, 97\%)\\ (91\%, 91\%)\end{tabular} \\
\hline
Q3   & \begin{tabular}[c]{@{}l@{}}(88\%, 90\%)\\ (87\%, 87\%)\end{tabular} & \begin{tabular}[c]{@{}l@{}}(92\%, 93\%)\\ (88\%, 88\%)\end{tabular} & \begin{tabular}[c]{@{}l@{}}(95\%, 96\%)\\ (88\%, 88\%)\end{tabular} \\
\hline
Q4   & \begin{tabular}[c]{@{}l@{}}(54\%, 55\%)\\ (60\%, 60\%)\end{tabular} & \begin{tabular}[c]{@{}l@{}}(55\%, 56\%)\\ (60\%, 60\%)\end{tabular} & \begin{tabular}[c]{@{}l@{}}(56\%, 56\%)\\ (60\%, 60\%)\end{tabular} \\
\hline
Q5   & \begin{tabular}[c]{@{}l@{}}(55\%, 56\%)\\ (60\%, 60\%)\end{tabular} & \begin{tabular}[c]{@{}l@{}}(56,\% 57\%) \\ (60\%, 59\%)\end{tabular} & \begin{tabular}[c]{@{}l@{}}(57\%, 57\%)\\ (59\%, 59\%)\end{tabular} \\
\bottomrule
\end{tabular}}
\caption*{ \footnotesize {\bf Note:} Each entry corresponds to the mean and median of pensions under the current system and under English auctions, expressed as a percentage of the pension under full information, separated by savings quintile (rows) and the number of potential bidders (columns). Each quintile has two rows: The first row corresponds to the current system and the second row corresponds to the English auction.}
\end{center}
\end{table}

Next, we also consider the money's worth ratio for each group. The money's worth ratio is the ratio of the discounted stream of payments associated with the annuity contract to the premium. So it is a measure of the ``generosity'' of the annuity contract.
If this ratio is more/less than one, we say a retiree expects to earn more/less than what she annuitizes. The money's worth ratio for $i$, under the current system, is the ratio of $i$'s chosen pension times $UNC_i$ to $S_i$.

In Table \ref{table:mwr_}, we present group-specific money's worth ratios, which are equal to the ratio $(\sum_{i} P_i\times UNC_i)/\sum_{i} S_i$, where the sum is over all retirees in the respective group. 
As we can see from the first column, those with AFP (the first row within each quintile) get better money's worth ratio than the other two channels under the current system. 
We also see that those with high savings get slightly better offers than lower savers. 
Comparing the first and the last columns in Table \ref{table:mwr_}, we see that as before, the gap between the current system and complete information is largest for high savers, which is consistent with the costs in Figure \ref{fig:cf_cdf_pensions}. 

\begin{table}[ht!]
\begin{center}
\caption{\bf Money's Worth Ratio, by Savings Quintile and Channel \label{table:mwr_}}
\scalebox{0.85}{\begin{tabular}{cccccc}
\toprule
{\bf Savings Quintile} & {\bf Channel}&{\bf Current} & {\bf English} & {\bf Full Info.}\\
\toprule
Q1&AFP&0.99018 & 0.93229 & 1.04419 \\
& Sales Agent&0.95663 & 0.93128 & 1.04327 \\
&Advisor&0.95969 & 0.93019 & 1.04237 \\
\hline
Q2&AFP&1.02480 & 0.95833 & 1.04920 \\
& Sales Agent&0.99589 & 0.95728 & 1.04841 \\
&Advisor&0.99624 & 0.95608 & 1.04748 \\
\hline
Q3&AFP&1.04418 & 0.96340& 1.08998 \\
& Sales Agent&1.02315 & 0.96216 & 1.08906 \\
&Advisor&1.01623 & 0.96067 & 1.08796 \\
\hline
Q4&AFP&1.06109 & 1.13492 & 1.86677 \\
& Sales Agent&1.04144 & 1.13166 & 1.86129 \\
&Advisor&1.03278 & 1.12759 & 1.85429 \\
\hline
Q5&AFP&1.09793 & 1.12368 & 1.87748 \\
& Sales Agent&1.07350 & 1.12027 & 1.87109 \\
&Advisor&1.06609 & 1.11688 & 1.86514 \\
\bottomrule
\end{tabular}}
\caption*{\footnotesize {\bf Note:} Each row denotes a different group, and each entry is money's worth ratio $(\sum_i P_i\times UNC_i)/\sum_{i} S_i$, where the sum is taken over all retirees in the group. There are 15 groups based on 5 savings quintiles and 3 channels. Each column corresponds to a different pricing mechanism, where English is the English auction.}
\end{center}
\end{table}

\subsection{English Auction}
One way to increase pensions is to make the system more competitive.
To this end, we can replace the current system with the standard English auction and also ``shut down" risk ratings on the supply side by picking the winner to be the firm that offers the highest pension.
Simplifying the process should improve outcomes for those who choose in the first round. 
Similarly, shutting down risk rating should force firms to bid more aggressively; the benefits should be more considerable for lower savers than higher savers. The former have stronger preferences for risk ratings, which means without risk ratings, the firms should be more aggressive if the retiree is of lower savings. 
However, because the gap between the chosen pension and the full information pension is the largest for those with higher savings, they may benefit the most from the new mechanism.

We implement standard English auctions by treating potential bidders as the actual bidders.
Our results are an upper bound on the effect of English auction on pensions and retirees' ex-post expected present discounted gross utilities.
We follow the same steps as in the complete information counterfactual, except that under the English auction, the winning pension is the maximum pension a firm with the second-lowest-cost can offer, at zero profit.  

We present the kernel density estimate of the distributions of winning pensions under English auction in Figure \ref{fig:cf_cdf_pensions}.
Although English auction leads to higher pensions, most benefits accrue to those in the top two savings quintiles. 
We can also see this in the second row of Table \ref{table:pension_full_info} for each quintile, where we present the mean and the median pension under English auction expressed as a percentage of the pension under complete information. 
Similar results hold if we consider the money's worth ratio; see the first two columns in Table \ref{table:mwr_}.

We are also interested in determining the effect of using the English auction on retirees' ex-post utilities. 
We do not know the utility from the outside option, but we can determine the ex-post gross expected discounted present utility, which is equal to $\beta_i \times Z_j + \rho_{ij} + \theta_i b_{ij}$. 

For each retiree and each mechanism using the ``winning" pensions, we first determine the bequest (if any) and then calculate the ex-post expected present discounted utilities. 
To shed light on the effect of shutting down risk ratings on retirees' utilities, we calculate two utilities for each mechanism: one with $\beta_i \times Z_j$ and one without $Z_j$ by setting $\beta_i=0$. 
To calculate the utility from the risk rating, we use simulated data under the assumption that $\beta_i$ is a Normal random variable, with an estimated group-specific mean and variance. 

We present the average utilities across different groups in Tables \ref{table:utility1} and \ref{table:utility2}. In Table \ref{table:utility1}, we group retirees by their savings quintiles and the potential number of bidders, and in Table \ref{table:utility2}, we group retirees by their savings quintiles and their channels. 
In each table, and for each mechanism, we have two columns, one with and one without (asterisk) $\beta$, respectively.

Note that for each quintile in Table \ref{table:utility1}, by comparing the rows, we can see that the utilities increase with the number of bidders because the pensions increase when there are more firms.
However, despite the large gap between the pensions under the current system or the pensions under English auction and the pension under the complete information (Figure \ref{fig:cf_cdf_pensions}), our estimates show that the gap in utilities is almost negligible. 
In Table \ref{table:utility2} we can see that similar results hold even if we group retirees by their savings quintile and channel. 
Nonetheless, what is new here is that those who have sale agents (second row in each savings quintile) have higher utilities than other channels, and the gap decreases with savings.\footnote{ Adding an optimal reserve price has an insignificant effect on the outcomes.}

\begin{table}[th!]
\begin{center}
\caption{ {\bf Average Gross Utility, by Savings Quintile and Potential Bidders \label{table:utility1}}}
\scalebox{0.85}{\begin{tabular}{cllllllll}
\toprule
{\bf Bidders} &{\bf Current} & {\bf English} &{\bf Full Info.} & {\bf Current}* & {\bf English}* & {\bf Full Info.}* \\
\toprule
13&8.8176 & 8.8180 & 8.8191 & -0.0054 & -0.0049 & -0.0039 \\
14&6.9852 & 6.9851 & 6.9866 & -0.0073 & -0.0073 & -0.0058 \\
15&11.9204 & 11.9200 & 11.9215 & -0.0073 & -0.0077 & -0.0061 \\
\hline
13&3.5616 & 3.5618 & 3.5622 & -0.0027 & -0.0026 & -0.0021 \\
14&3.5055 & 3.5054 & 3.5061 & -0.0038 & -0.0040 & -0.0033 \\
15&4.2757 & 4.2753 & 4.2760 & -0.0042 & -0.0046 & -0.0039 \\
\hline
13&2.5903 & 2.5903 & 2.5907 & -0.0015 & -0.0014 & -0.0011 \\
14&2.6788 & 2.6787 & 2.6791 & -0.0018 & -0.0020 & -0.0015 \\
15&2.8087 & 2.8084 & 2.8089 & -0.0021 & -0.0024 & -0.0018 \\
\hline
13&2.4089 & 2.4091 & 2.4095 & -0.0007 & -0.0005 & -0.0002 \\
14&2.4462 & 2.4464 & 2.4468 & -0.0009 & -0.0007 & -0.0003 \\
15&2.4724 & 2.4726 & 2.4731 & -0.0010 & -0.0008 & -0.0003 \\
\hline
13&2.3357 & 2.3358 & 2.3359 & -0.0003 & -0.0002 & -0.0001 \\
14&2.2684 & 2.2684 & 2.2686 & -0.0004 & -0.0003 & -0.0001 \\
15&2.3018 & 2.3019 & 2.3021 & -0.0004 & -0.0004 & -0.0001\\
\bottomrule
\end{tabular}}
\caption*{\footnotesize {\bf Note:} The table displays the gross utility (\ref{eq:utility}), under current, English auction, and full information pricing mechanisms, averaged over groups defined by savings quintile and potential bidders. Quintiles are separated by horizontal lines, and within each quintile, the rows are the number of potential bidders $\{13,14,15\}$. The first four columns use estimated $\beta$ (Figure \ref{fig:beta}) for utilities and the last four (with asterisk) set $\beta=0$ in (\ref{eq:utility}). }
\end{center}
\end{table}

\begin{table}[ht!]
\begin{center}
\caption{ {\bf Average Gross Utility, by Savings Quintile and Channel \label{table:utility2}}}
\scalebox{0.85}{\begin{tabular}{cllllllll}
\toprule
{\bf Channel} &{\bf Current} & {\bf English} &{\bf Full Info.} & {\bf Current}* & {\bf English}* & {\bf Full Info.}* \\
\toprule
AFP&9.2078 & 9.2073 & 9.2087 & -0.0066 & -0.0071 & -0.0057 \\
Sales Agent&11.7779 & 11.7778 & 11.7794 & -0.0075 & -0.0075 & -0.006 \\
Advisor&9.239 & 9.2388 & 9.2402 & -0.0068 & -0.0069 & -0.0055 \\
\hline
AFP&3.7995 & 3.799 & 3.7998 & -0.0038 & -0.0043 & -0.0036 \\
Sales Agent&4.4095 & 4.4092 & 4.4099 & -0.0041 & -0.0043 & -0.0036 \\
Advisor&3.585 & 3.5848 & 3.5854 & -0.0038 & -0.004 & -0.0033 \\
\hline
AFP&2.6741 & 2.6738 & 2.6743 & -0.0019 & -0.0022 & -0.0017 \\
Sales Agent&2.9609 & 2.9607 & 2.9611 & -0.002 & -0.0022 & -0.0017 \\
Advisor&2.5351 & 2.535 & 2.5354 & -0.0019 & -0.0021 & -0.0016 \\
\hline
AFP&2.4637 & 2.4639 & 2.4644 & -0.0009 & -0.0008 & -0.0003 \\
Sales Agent&2.5845 & 2.5847 & 2.5852 & -0.001 & -0.0008 & -0.0003 \\
Advisor&2.2824 & 2.2826 & 2.283 & -0.0009 & -0.0007 & -0.0003 \\
\hline
AFP&2.3075 & 2.3076 & 2.3078 & -0.0004 & -0.0003 & -0.0001 \\
Sales Agent&2.3537 & 2.3537 & 2.354 & -0.0004 & -0.0004 & -0.0001 \\
Advisor&2.2215 & 2.2216 & 2.2218 & -0.0004 & -0.0003 & -0.0001\\\bottomrule
\end{tabular}}
\caption*{\footnotesize {\bf Note:} The table displays the gross utility, (\ref{eq:utility}), under current, English auction, and full information pricing mechanisms, averaged over groups defined by savings quintiles and channels (AFP, sales agents and advisors). The first four columns use the estimated $\beta$ (Figure \ref{fig:beta}) and the last four (with asterisk) set $\beta=0$ in (\ref{eq:utility}). }
\end{center}
\end{table}

\section{Conclusion \label{section:conclusion}}
In this paper, we develop an empirical framework to study an imperfectly competitive market for annuities. 
We used a rich administrative data set from the Chilean annuity market to estimate our model. 
In the market, risk-averse retirees use two-stage multi-attribute auctions to select from different types of annuity contracts and different firms. 
Life insurance companies have private information about their annuitization costs, and for each retiree auction, they decide whether to participate and compete by making pension offers. 
The Chilean data give us a unique opportunity to examine the role of private information about cost, retiree's preferences, and market structure on the outcomes of a market for annuities. 

Our main contribution is to study the current market system by estimating the demand and supply of annuities and evaluating a simpler mechanism that may improve the system. 
While there is a gap between the observed pensions under the current system and pensions under the complete information regime, the gap is significantly larger for those with higher savings. We also determine the effect of replacing the current system with a simpler one-shot English auction, where the winning firm offers the highest pension on pensions and ex-post expected present discounted utilities. 
We find that while the new mechanism increases pensions for almost every retiree, pensions increase the most for those in the top two savings quintiles, albeit the increase in utility is minimal. 

A possible avenue for future research is to include the choice between PW and annuities and consider an imperfectly competitive market with two-sided asymmetric information. 
On the demand side, retirees may have private information about their mortality forces and their bequest preferences. On the supply side, as in our case, firms have private information about their annuitization costs. Such a model would shed light on the interaction of competition and adverse selection and thus provide a richer insight into the annuity market than our current framework.

\bibliographystyle{jpe}
\bibliography{annuity}

\clearpage 
\newpage

\setcounter{section}{0}
\setcounter{equation}{0}
\setcounter{table}{0}
\setcounter{figure}{0}
\renewcommand{\thesection}{ A}
\renewcommand{\theequation}{A.\arabic{equation}}
\renewcommand{\thetable}{A.\arabic{table}}
\renewcommand{\thefigure}{A.\arabic{figure}}
\begin{center}
\large{\bf Appendix}
\end{center}
{\footnotesize

\section{Intermediary Channels\label{section:channels}}
We observe retirees with one of the four intermediary channels (AFP, insurance company, sales agent, or independent advisor) to assist them with their annuitization process. However, they may also ``influence" the final decision, especially if the sales agent gets paid only if the retiree chooses the agent's firm. One way for an agent to influence the decision would be to emphasize the importance of risk ratings to retirees. In our empirical setting, we allow preferences for risk ratings and information processing costs to depend on the channel to capture this effect. 

To understand if intermediaries systematically affect pensions, we compare channel-specific money's worth ratio (henceforth, \mwr), which is the expected present value of pension per annuitized dollar. The {\mwr} measures the generosity of an annuity contract \citep{MitchellPoterbaWarshawskyBrown1999}.\footnote{ If ${\mwr} =1$ then retiree expects to get \$1 (in present discounted value) for every annuitized dollar.}
In Figure \ref{fig:res_all2}, we display the distributions of the {\mwr} offered in the first round (left panel) and {\mwr} accepted by the retirees (right panel). 
 The mean and the median {\mwr} of the offers, by channels (AFP, insurance company, sales agent, advisor), are $(0.989, 0.988, 0.984, 0.987)$ and $(0.990, 0.989, 0.986, 0.988)$, respectively, but the means and medians for accepted offers are $(1.010, 1.010, 0.990, 1.007)$ and $(1.010,1.009, 0.991, 1.007)$, respectively. Thus, the final accepted offers are on average better than the first-round offers, and those with sales agents have slightly lower {\mwr}.

\begin{figure}[h!]
\caption{\bf CDFs of Offered and Accepted MWR, by Channel}
\hspace{-0.5in}
\includegraphics[scale=0.3]{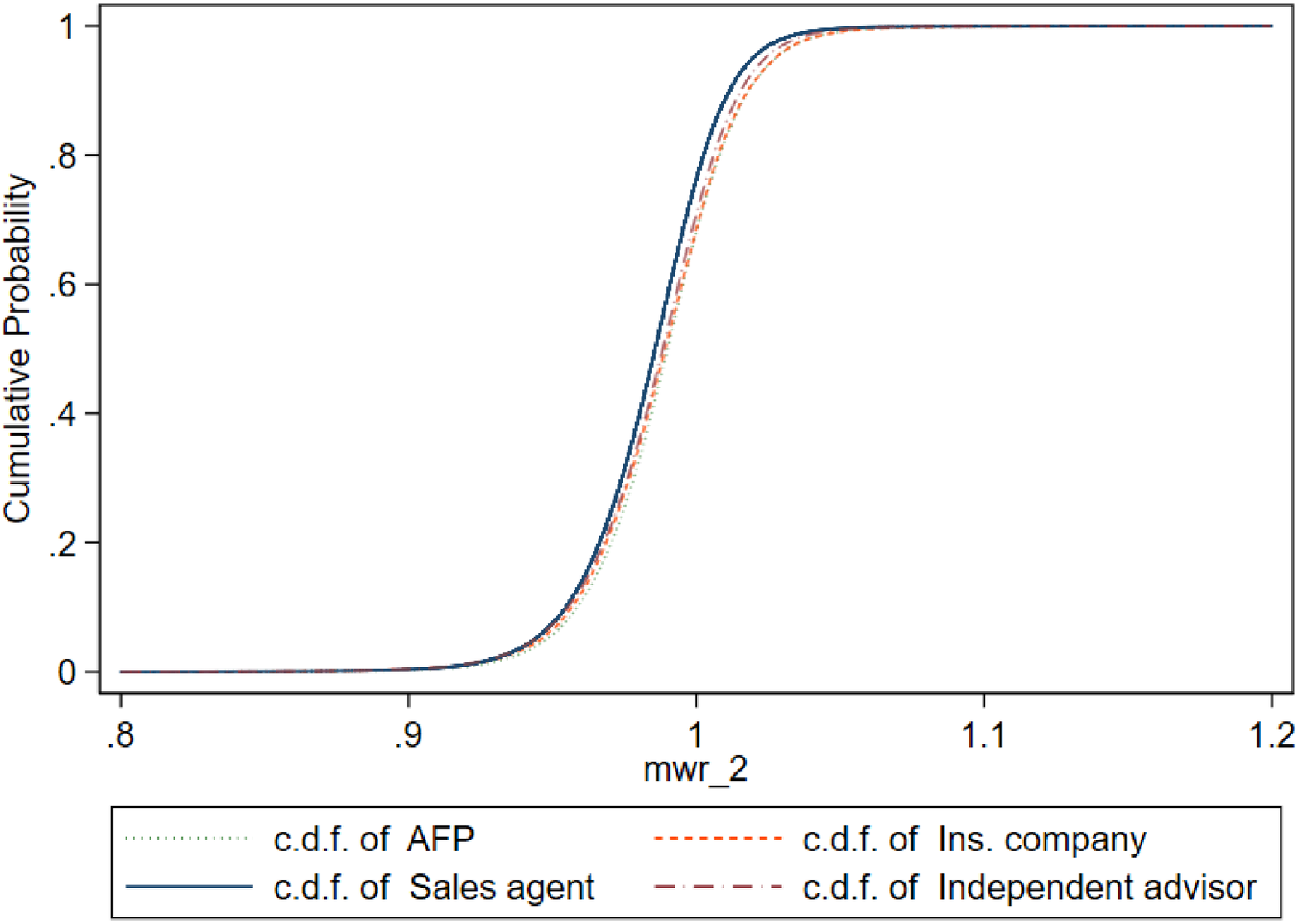}~
\includegraphics[scale=0.3]{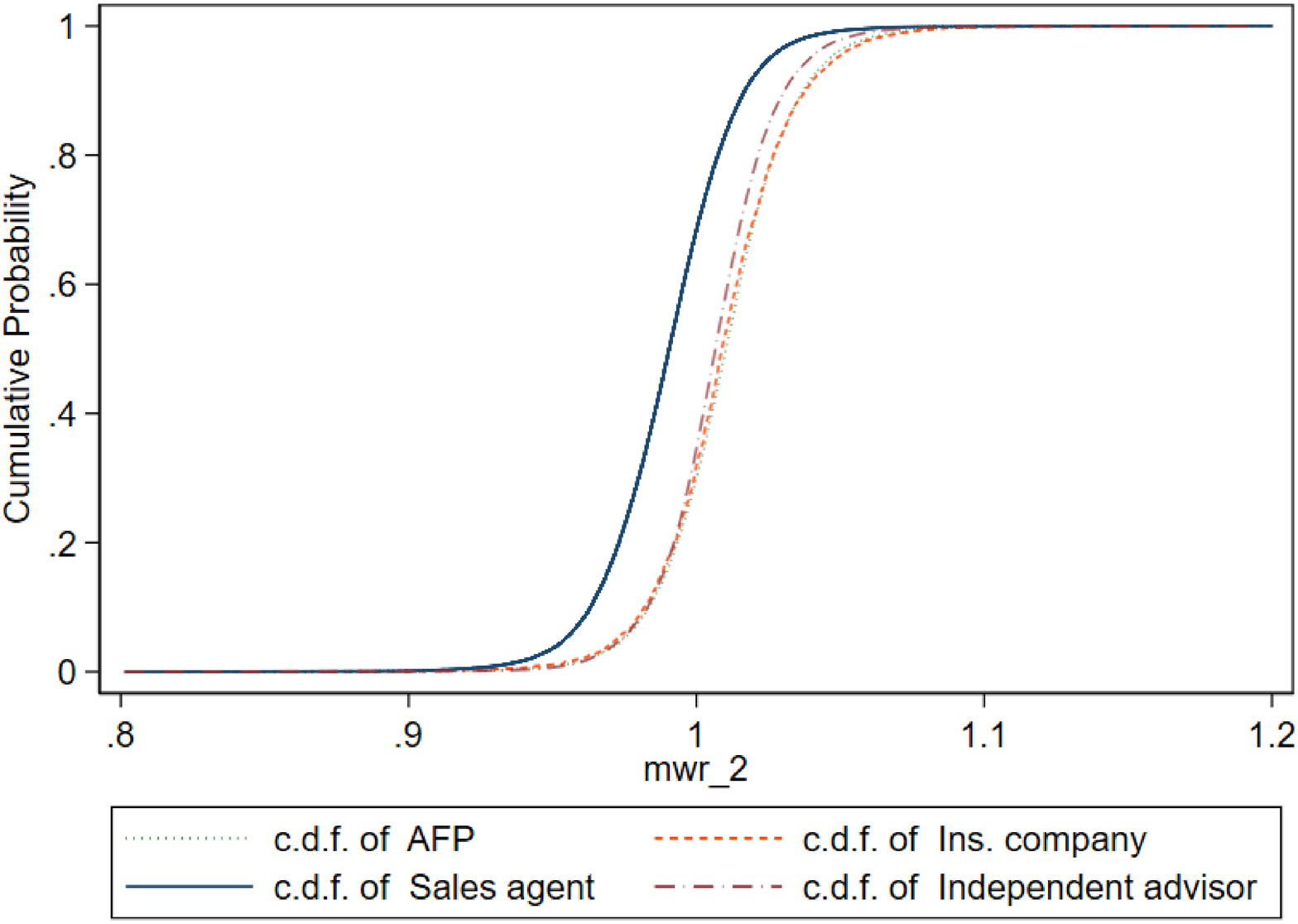}
\label{fig:res_all2}
\caption*{\footnotesize {\bf Note:} Distributions of the offered and chosen {\mwr} (left panel vs. right panel), by channel.}
\end{figure}

More formally, we use a multinomial logit regression model to verify if observed differences among retirees can explain the differences in their channels; see Table \ref{tab:mlogit}.  
We see that some characteristics and the channel are correlated. For instance, those who have lower savings, retire early, are male, or are unmarried are more likely to use sales agents than AFP.

\begin{table}[th!]\centering \caption{\bf Intermediary Channel - Estimates from Multinomial Logit \label{tab:mlogit}}
\scalebox{0.8}{\begin{tabular}{lcccc}
\toprule
 {\bf Regressors $\backslash$ Channels}& {\bf Insurance Company} & {\bf Sales-Agent} & {\bf Advisor} \\ \toprule
 Savings (\$million)   & 0.629*** & -0.857*** & -0.130*** \\
   & (0.128) & (0.0436) & (0.0447) \\
 Age   & 0.0131 & -0.0408*** & -0.0816*** \\
 &   (0.00857) & (0.00189) & (0.00218) \\
 Female   & 0.437*** & -0.0588*** & -0.124*** \\
 &   (0.0546) & (0.0120) & (0.0140) \\
 Married   & 0.0245 & 0.0620*** & 0.0874*** \\
 &   (0.0491) & (0.0107) & (0.0127) \\
 Constant   & -5.029*** & 2.333*** & 4.326*** \\
 &   (0.560) & (0.123) & (0.142) \\
  \midrule
 N & 238,548 & 238,548 & 238,548 \\ \bottomrule
 \end{tabular}}

\caption*{\footnotesize {\bf Note}. Estimates of multinomial logit regression for channels, where the baseline choice is AFP. Standard errors are in parentheses, and $^{***}, ^{**}, ^{*}$ denote p-values less than $0.01, 0.05$ and $0.1$, respectively. } 
\end{table}

In the paper, we treat the intermediary channel as exogenous. There are at least three reasons why we believe this is not as strong an assumption in our context as it might appear. First, several anecdotal pieces of evidence suggest that most people rely on word-of-mouth when it comes to a channel suggesting that the observed channel-retiree ``match" is somewhat exogenous. Second, when making first-round offers, firms do not know retirees' channel choices, and it is reasonable to assume that the channel does not affect the offers. 

Third, because a channel, especially a sales agent, is supposed to help retirees ``process information," to capture this dependence, in our empirical application, we allow information-processing costs and preferences for risk ratings to depend on the channel by estimating these parameters separately for each group where a group is a combination of age, gender, savings, and channel. In particular, we posit that channels affect the cost of acquiring information about the importance of risk rating. For instance, we allow those retirees who use sales agents to act ``as if" they have a higher cost of acquiring information about the trade-off between risk rating and pensions. We assume that in the first stage, retirees are rationally inattentive with respect to their preference for risk ratings, but they know their preferences in the second stage.  

\setcounter{section}{0}
\setcounter{equation}{0}
\setcounter{table}{0}
\setcounter{figure}{0}
\renewcommand{\thesection}{ B}
\renewcommand{\theequation}{B.\arabic{equation}}
\renewcommand{\thetable}{B.\arabic{table}}
\renewcommand{\thefigure}{B.\arabic{figure}}
\newpage
\section{Additional Figures}
\begin{figure}[ht!]\caption{\bf Example of a Certificate of Offers\label{fig:scomp_picture}}
\hspace{-0.5in}
 \fbox{\includegraphics[scale=0.3]{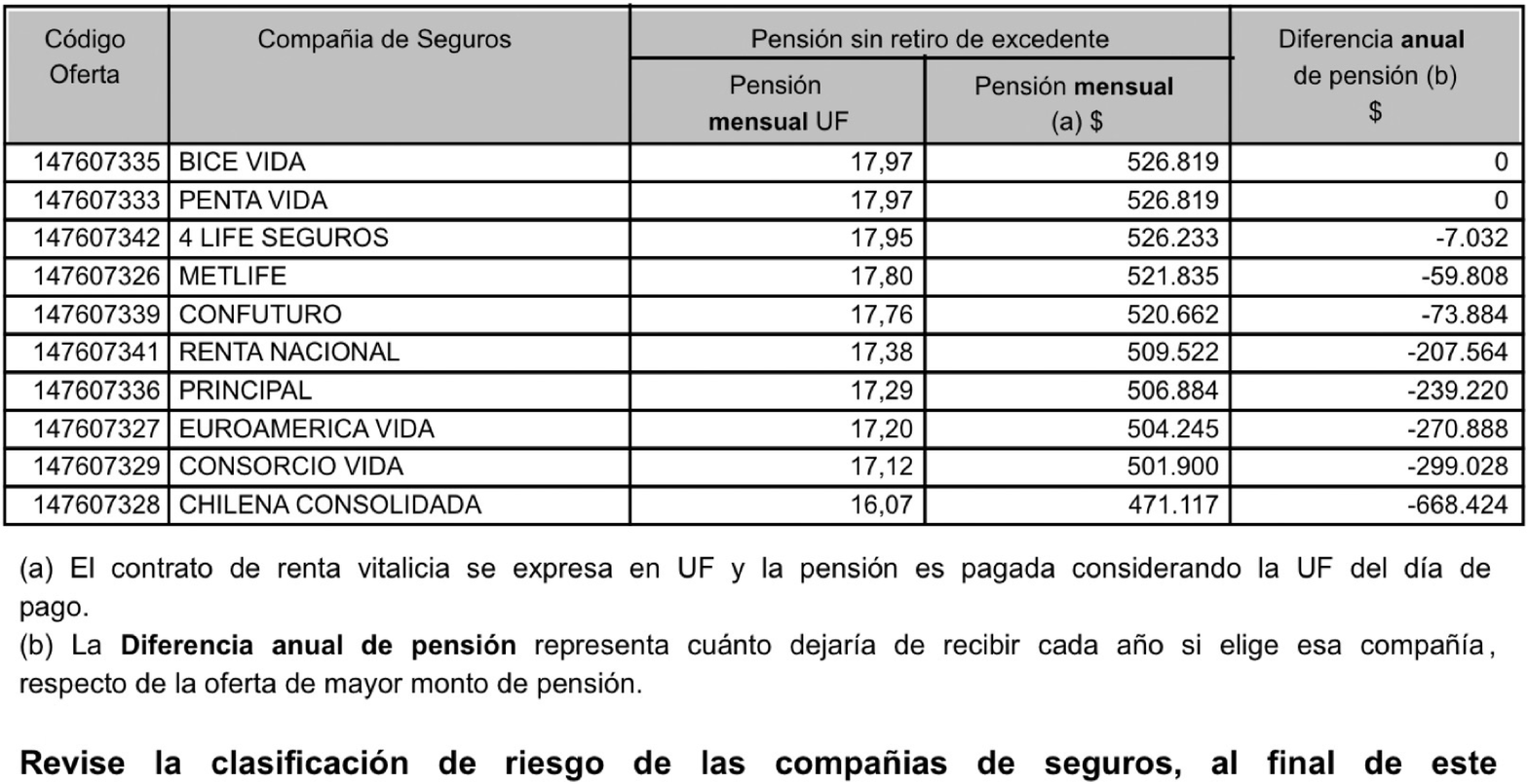}}  
\caption*{\footnotesize {\bf Note:} This is an example of \emph{Certificate of Offers} for an immediate annuity- it is a document that the SCOMP sends to the retiree if immediate annuity is one of the types of annuity contracts the retiree solicits offers for. The document is presented as a table, where the first column is the Offer Code; the second column is the name of the life insurance company ordered in terms of offered pensions, the third column shows the monthly pensions, first expressed in the \emph{Unidad de Fomento} (UF), which is a unit of account used in Chile, and the second expressed Chilean pesos (CLP\$); and the last column shows the amount the retiree \emph{would not} receive in a year if she chooses that company compared to the highest pension. For instance, if the retiree chooses the third firm (4 Life Seguros), every year she would lose CLP\$ 7,032 compared to the first firm (Bice Vida). There are similar, but separate, tables for different types of annuities. See \url{https://www.dropbox.com/s/vrfch6pcd91u604/longercertofquotes.pdf?dl=0} for a translated example of a \emph{Certificate of Offers} that includes several products with different GP and DP. Notice that in the table above, a comma separate decimals and a period separates thousands.}
\end{figure} 
\begin{figure}[ht!]\caption{\bf Pension Rates and $UNC_i$, by Firm\label{fig:UNCi}}
\hspace{-1.4in}
 \includegraphics[scale=0.65]{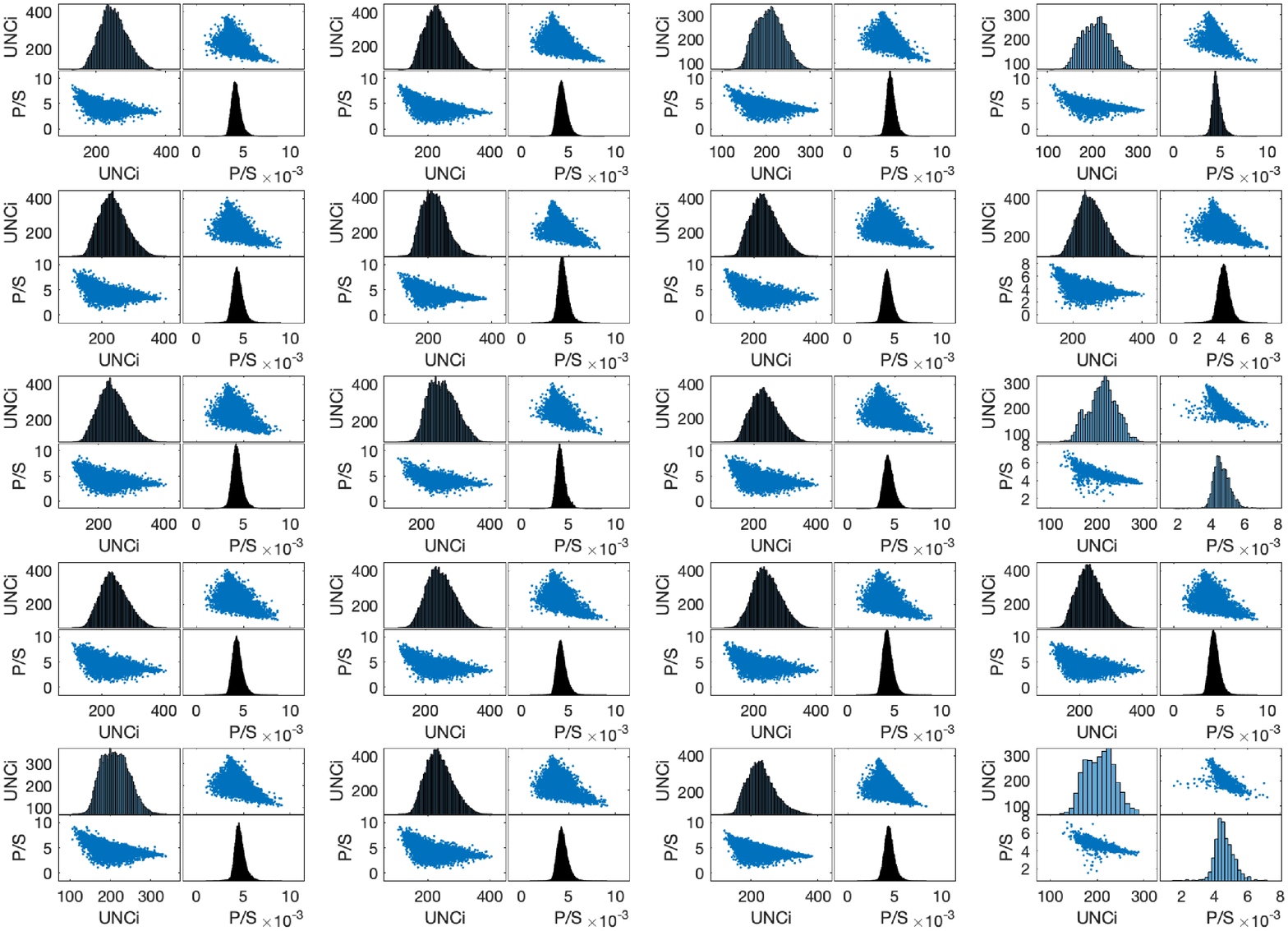}  
\caption*{\footnotesize {\bf Note:} These are histograms and scatter plots of monthly pension rates, i.e., the ratio of monthly pension to annuitized savings, and the $UNC_i$ of the retirees to whom the firms make an offer. There are 20 firms, so there are 20 sets of four subfigures each. Clockwise, the first subfigure is the histogram of $UNC_i$, and the second subfigure is the scatter plot of the pension rates (on the x-axis) and $UNC_i$ (on the y-axis). The third subfigure is the histogram of the pension rates, and the last subfigure is the scatter plot of $UNC_i$ and the pension rates with $UNC_{i}$ on the x-axis. Thus the second and fourth subfigures show exactly the same information but with axes that are flipped. }
\end{figure} 

\clearpage
\newpage
\setcounter{section}{0}
\setcounter{equation}{0}
\setcounter{table}{0}
\setcounter{figure}{0}
\renewcommand{\thesection}{ C}
\renewcommand{\theequation}{C.\arabic{equation}}
\renewcommand{\thetable}{C.\arabic{table}}
\renewcommand{\thefigure}{C.\arabic{figure}}

\section{ Present Expected Utilities\label{section:Gompertz}}

In this section, we explain how we determine the present expected utility (or Net Present Expected Value, NPEV) from an annuity contract with pension and bequest $(P_{ij}, B_{ij})$, i.e., $\mathfrak{U}(P_{ij}, B_{ij};\theta_i)$, under the assumption that retirees consume all their pension, every month. This process requires three steps. First, for each retiree, we determine the survival probability. Second, depending on the contract, we determine the (time-varying) bequest. Third, we calculate the expected present discounted utility using the average rate of return as the discount factor. 

\subsection{Basic Idea}
For intuition, consider only the first month after retirement, and let $q_{i}\in[0,1]$ be the probability that $i$ is alive one month after retirement. 
Then, the expected present discounted utility will be 
\begin{eqnarray*}
\mathfrak{U}(P_{ij},B_{ij}; \theta_i) = u(P_{ij}) \times q_{i}+ \theta_i \times v(B_{ij})\times (1-q_{i}), 
\end{eqnarray*}
where $u(P_{ij})$ is the flow utility from $P_{ij}$, and $v(B_{ij})$ is the utility from leaving a bequest $B_{ij}=B_{ij}(P_{ij})$. 
Now, let us consider two months after retirement. We have to adjust the probability that the retiree survives two periods given that she is alive at retirement and consider that the bequest left upon death will also change, which in turn depends on whether the annuity product under consideration includes a guaranteed period. \footnote{In practice, however, we do not observe for how long $i$ expects to live. 
So, to determine expected longevity at retirement, we estimate a continuous-time Gompertz survival function for $i$ and her spouse (if she is married) as a function of her demographic and socioeconomic characteristics. We explain the estimation method in the following subsection. We discuss the estimation later in this section.} 

Once we have the survival probabilities, the expected discounted utilities become the product of $u(P_{ij})$ and the discounted number of months $i$ expects to live. 
When there is bequest, then $\mathfrak{U}(P_{ij},B_{ij}(P_{ij}); \theta_i)$ has an intuitive structure. It is a sum of two terms, one of which is the product of $u(P_{ij})$ and the discounted number of months $i$ expects to live, and the other term is the product of $v(B_{ij})$ times the discounted number of months $i$'s beneficiaries expect to receive $B_{ij}$. 

Legally, $i$'s spouse is entitled to $60\%$ of $i$'s pension and $100\%$ during the guaranteed period; the amount $B_{ij}$ may change over time. 
Thus, we can write $\mathfrak{U}(P_{ij},B_{ij}(P_{ij}); \theta_i)$ as 
\begin{eqnarray}
\mathfrak{U}(P_{ij},B_{ij}(P_{ij}); \theta_i) &:=&u(P_{ij})\times D_i^R +\theta_i \left(v(0.6\times P_{ij}) \times D_i^{S} + v(P_{ij}) \times D_i^{S,GP} \right) \notag\\
&\equiv&\rho_i(P_{ij}) +\theta_i \times b_i(P_{ij}), \label{eq:utility11}
\end{eqnarray}
where $D_i^R$ is the discounted expected longevity of the retiree (in months, from the moment the annuity payments start) and $D_i^{S,GP}$ is the discounted number of months that the spouse (or other beneficiaries) will receive the full pension because of the guaranteed period. Furthermore, $D_i^{S}$ is the discounted number of months that the spouse will receive $60\%$ of the retiree's pension.\footnote{ These ``discounted life expectancies'' also can be interpreted in terms of the annuitization costs. Assuming firms use the same mortality process as we do and invest retirees' savings at an interest rate equal to the discount rate, then $D_i^R$ is the capital necessary to provide a one-dollar pension to the retiree until she dies. Similarly, $D_i^{S,GP}$ is the necessary capital to finance a dollar of pension for the beneficiaries once the retiree is dead and until the guaranteed period expires. Finally, $D_i^{S}$ is the necessary capital to finance a dollar of pension for the beneficiaries between the retiree's death or the guaranteed period is over (whichever occurs later) and until the spouse dies. The gains from trade between retirees and insurance companies come from the differences in risk-attitude between retirees and life insurance companies and potential differences between retirees' discount rates and firms' investment opportunities. } 
If the annuity has a deferred period, the retiree gets twice her pension until the annuity payment begins. So $\rho_i(P_{ij})=u(P_{ij})\times D_i^R+ u(2 P_{ij})\times D_i^{R, DP}$ where $D_i^{R, DP}$ is the expected life during the deferred period.\footnote{For simplicity, we are disregarding survival benefits during the deferment period. Deferred periods in our sample are at most three years. Thus, the death probability is quite low.} 

Now suppose the annuity includes guaranteed periods, and we consider more than two months. 
When the annuity includes a guaranteed period (of $G$ months), and the annuitant dies before $G$, say in $G'<G$ months, her spouse will continue to get the same pension for the next $(G-G')$ months, and after that, he gets 60\% of the original pension. 
If there is no surviving spouse at the time of death, either because the retiree was single or widowed, the designated beneficiaries get 100\% of the pension. 
We assume that the retiree values her spouse or other beneficiaries in the same way, with utility $v(B_{it})$.
Using these rules (\ref{eq:utility11}) becomes
{\begin{eqnarray}
\mathfrak{U}(P_{ij},B_{ij}(P_{ij}); \theta_i) &=&u(P)\times D_i^R+ \theta_i \times \left(\sum_{t=0}^G \frac{(1-q_{it})}{(1+\delta_{t})^t} \times v(P) + \sum_{t=G+1}^T \frac{(1-q_{it})q_{it}^*}{(1+\delta_{t})^t} \times v(0.6\times P) \right)\notag\\
&=&u(P)\times D_i^R+ \theta_i \times \left(v(P)\sum_{t=0}^G \frac{(1-q_{it})}{(1+\delta_{t})^t} + v(0.6 \times P)\sum_{t=G+1}^T \frac{(1-q_{it})\times q_{it}^*}{(1+\delta_{t})^t} \right)\notag\\
&=&u(P)\times D_i^R+\theta_i \times \left(v(P) \times D_i^{S} + v(0.6\times P) \times D_i^{S,GP} \right),\label{eq:w}
\end{eqnarray}}
where $q_{it}^*$ is the probability that the spouse will be alive in $t$.

\subsection{Calculation of Present Expected Utilities}
Next, we explain how we determine the NPEV of an annuity given by Equation (\ref{eq:utility11}). 
To provide intuition, while keeping the notations manageable, we explain only a simple case where mortality probabilities are known and common across all individuals. Once we understand this simpler case, it is straightforward to allow for individual-specific longevity prospects but notationally messy, and for brevity, we do not describe that case here. 

The main difficulty in determining Equation (\ref{eq:utility11}) is that, unlike a constant pension, bequest (the wealth left for her estate) varies over time and across retirees. In particular, it depends on having legal beneficiaries, the type of annuity (in particular, whether it has a guaranteed period), and the time of death (before or after the guaranteed period).
Chilean law states that specific individuals are eligible to receive survivorship benefits upon the death of a retiree. As mentioned in Section \ref{section:data}, we focus on retirees without eligible children (but with or without spouses), which is the most common case in our sample. The spouse is eligible for a survivorship annuity equivalent to 60\% of the retiree's original pension.

\noindent{\bf Probability of Death.} First, we model the force of mortality as a continuous random variable distributed as a Gompertz distribution.
Let $F_m(t|X)$ be the conditional distribution function for the time of death of retiree with characteristics $X$, and let $f_m(t|X)$ be the corresponding conditional density. For notational simplicity, we suppress the dependence on $X$. The probability of being alive at time $t$, i.e., that death occurs after $t$, is given by the survivor function $\overline{F}_m(t):=1-F_m(t)$. we assume that $F_m$ follows a Gompertz distribution, so the conditional survival functions as $\overline{F}_m(t\mid t>t_0; \lambda, \mathfrak{g})=e^{-\frac{\lambda}{\mathfrak{g}}(e^{\mathfrak{g} t}-e^{\mathfrak{g} t_0})}.$ 
Since the analysis is from the perspective of a retiree who is alive at $t_0$, henceforth, all relevant functions are conditional on being alive at $t_0$. To allow demographic characteristics $X$ to affect mortality, we let $\lambda=\exp\left(X^{\top}\tau\right)$, and estimate the parameters $(\mathfrak{g}, \tau)$ using maximum likelihood estimation; see Section \ref{section:mortality}.  

\noindent{\bf Immediate Annuity.} Let $t_0$ denote the age at retirement, expressed in months and let $\delta\in(0,1)$ denote the discount factor. 
An annuity pays a constant benefit $P$ from $t_0$ until retiree's death, so NPEV is calculated at $t_0$. 
We start by considering an immediate annuity with no spouse. Such annuity does not pay anything to the beneficiaries upon death; therefore, $b_{ij}=0$. 
Then the NPEV of an immediate annuity is 
\begin{equation}
\label{eq:NEPV0}
\rho=\int_{t_0}^{\infty}u(P)\overline{F}_m(t|t>t_0)e^{-\delta(t-t_0)}dt.
\end{equation}
As introduced in Section \ref{section:mortality}, we assume that $F_m$ is a Gompertz distribution, so the conditional survival functions as $\overline{F}_m(t\mid t>t_0; \lambda, \mathfrak{g})=e^{-\frac{\lambda}{\mathfrak{g}}(e^{\mathfrak{g} t}-e^{\mathfrak{g} t_0})}.$
Substituting $\overline{F}_m(t\mid t>t_0; \lambda, \mathfrak{g})=e^{-\frac{\lambda}{\mathfrak{g}}(e^{\mathfrak{g} t}-e^{\mathfrak{g} t_0})}$ in (\ref{eq:NEPV0}) gives 
\begin{equation}
\label{eq:NPEV1}
\rho=u(P)\times \left\{e^{\delta t_0}e^{\frac{\lambda}{\mathfrak{g}}e^{\mathfrak{g} t_0}}\int_{t_0}^{\infty}e^{-\frac{\lambda}{\mathfrak{g}}e^{\mathfrak{g} t}}e^{-\delta t}dt\right\}=u(P) \times D^R,
\end{equation}
where recall that $u(P)=\frac{P^{(1-\gamma)}}{1-\gamma}$ with risk aversion parameter $\gamma=3$ and the discount factor $\delta=\ln(1+\tilde{r}_{t_0})$, with $\tilde{r}_{t_0}$ is the annual market rate of return at $t_0$.

{\bf Deferred Annuity.} If the annuity contracts include a deferred period clause for $d$ months, then the pensions start from $t_0+d$. 
In the meantime, the retiree receives a ``temporal payment," which is almost always twice the pension.
The annuity component of the NPEV expression in (\ref{eq:NPEV1}) remains the same, except the lower limit is $t_0+d$ and an additional term reflecting the temporal payment to be received during the transitory period:
\begin{equation}
\label{eq:NPEV2}
\rho= u(2P)\times \left\{e^{\delta t_0}e^{\frac{\lambda}{\mathfrak{g}}e^{\mathfrak{g} t_0}}\int_{t_0}^{t_0+d}e^{-\frac{\lambda}{\mathfrak{g}}e^{\mathfrak{g} t}}e^{-\delta t}dt\right\}+u(P)\times \left\{e^{\delta t_0}e^{\frac{\lambda}{\mathfrak{g}}e^{\mathfrak{g} t_0}}\int_{t_0+d}^{\infty}e^{-\frac{\lambda}{\mathfrak{g}}e^{\mathfrak{g} t}}e^{-\delta t}dt\right\}.
\end{equation}
 
{\bf Annuity with Guaranteed Periods.} 
Besides deferment, annuity contracts can also have a guaranteed period clause, which implies that if the retiree dies within a certain period (denoted as $g$ months) from the start of the payment (either $t_0$ or $t_1=t_0+d$), the total pension amount ($P$) will be paid to the retiree's spouse or other beneficiaries specified in the contract until the end of the guaranteed period.
The NPEV of benefits to be received by the retiree is the same as (\ref{eq:NPEV1}) if $d=0$ and (\ref{eq:NPEV2}) if $d>0$.
As the retiree's beneficiaries are now eligible for benefits in the event of death within the guaranteed period, we let $b$ as the NPEV of benefits received by these beneficiaries, i.e., bequests. 
Recall that the instantaneous utility associated with beneficiaries receiving a pension $P$ is given by $\theta \times v(\cdot)$, where, recall that $v(B)=\frac{B^{(1-\gamma)}}{1-\gamma}$ with risk-aversion parameter $\gamma=3$.

The bequest $b$, assuming a deferment period until $t_0+d$ and a guaranteed period of $g$, is similar to (\ref{eq:NEPV0}), except that the upper integration limit is given by the guaranteed period and the instantaneous probability function corresponds to $F_m(t\mid t>t_0; \lambda, \mathfrak{g})$:

 \begin{eqnarray}
 \label{eq:NPEV3}
b= v(P)\times\left\{\int_{t_1}^{t_1+g}\!\!\!\!\!\!F_m(t|t>t_0; \lambda, \mathfrak{g})e^{-\delta(t-t_0)}dt\right\}= v(P)\times \left\{ \int_{t_0+d}^{t_0+d+g}\!\! \!\!\!\!(1-e^{-\frac{\lambda}{\mathfrak{g}}(e^{\mathfrak{g} t}-e^{\mathfrak{g} t_0})})e^{-\delta(t-t_0)}dt\right\}.\quad
 \end{eqnarray}

{\bf Allowing for Eligible Spouse.}
When a participant is married at the time of retirement, the spouse is eligible for a survivorship benefit if he or she outlives the retiree. This benefit is until death and, in the absence of eligible children, equivalent to 60\% of the original pension benefit.
Once again, the formula for NPEV associated with benefits to be received by the retiree ($\rho$) is not affected by the spouse's presence (except that the offered pension will be lower to account for the additional contractual entitlements). 

The formula for the NPEV of bequest must then include an additional term, to account for the additional benefits to be paid in the case the spouse outlives the retiree, after the guaranteed period has elapsed. We assume that the two mortality processes are independent and follow the same Gompertz distribution (same $\mathfrak{g}$ parameter, but different $\lambda_{sp}$ parameter for the spouse). In this case, the expression for NPEV of bequest is given by:
 \begin{eqnarray}
\label{eq:NPEV4}
b&=&\int_{t_0+d}^{t_0+d+g} v(P)\times F_m(t|t>t_0; \lambda, \mathfrak{g})\times e^{-\delta(t-t_0)}dt\notag\\&& +\int_{t_0+d+g}^{\infty}\!\!\! v(0.6\times P)\times F_m(t|t>t_0; \lambda, \mathfrak{g})\times \overline{F}_m(t-\Delta|t-\Delta>t_0-\Delta;\lambda_{sp}, \mathfrak{g})\times e^{-\delta(t-t_0)}dt \notag\\
&=& v(P)\int_{t_0+d}^{t_0+d+g}(1-e^{-\frac{\lambda}{\mathfrak{g}}(e^{\mathfrak{g} t}-e^{\mathfrak{g} t_0})})\times e^{-\delta(t-t_0)}dt \notag\\ &&+ v(0.6\times P)\times \int_{t_0+d+g}^{\infty}(1-e^{-\frac{\lambda}{\mathfrak{g}}(e^{\mathfrak{g} t}-e^{\mathfrak{g} t_0})})\times(e^{-\frac{\lambda_S}{\mathfrak{g}}(e^{\mathfrak{g} (t-\Delta)}-e^{\mathfrak{g} (t_0-\Delta)})})\times e^{-\delta(t-t_0)}dt,
\end{eqnarray}
where $\Delta$ is the age difference between the retiree and the retiree's spouse.

\subsection{Mortality\label{section:mortality}}
A determinant of annuity demand and supply is the retiree's expected life or longevity. 
For a retiree in our sample, we observe her retirement age and her age at death if she dies by December 2017. 
To predict expected age at death, given information at retirement, we estimate a proportional hazard model. 

Let $h_{it}$ be the hazard rate for retiree $i$ with socioeconomic characteristics $X_i$ at time $t \in\mathbb{R}_+$, that includes $i$'s age, gender, marital status, savings and the year of birth, i.e., $h_{it}=\lim_{dt\rightarrow0}\frac{d\Pr(m_{i}\in[t,dt)|X_i, m_i\geq t)}{dt}=h(X_{i})\times \psi(t)$,
where $m_i$ is $i$'s realized mortality date, $\psi(t)$ is the baseline hazard rate given by a Gompertz distribution, such that the probability of $i$'s death by time $t$ is $F_m(t; \lambda_i, \mathfrak{g})=1-\exp(-\frac{\lambda_i}{\mathfrak{g}}\left(\exp(\mathfrak{g} t)-1)\right)$, with $\lambda_i = \exp(X_i^{\top}\tau)$. 
The identification of such a model is well established in the literature \citep{vanderberg2001}.
 
 Maximum likelihood estimates of the Gompertz suggest a smaller hazard risk is associated with younger cohorts, individuals who retire later, with females, those who are married, and those with higher savings.\footnote{ For robustness, we consider a new dataset that contains mortality information from before SCOMP (2004). The estimates are qualitatively the same. For instance, the predicted median expected life at death is 85 and 96 for males and females, respectively. The estimates are available from the authors upon request.} The implied median expected lives, by gender and savings quintile, and their standard errors are in Table \ref{table:mediantime}. 
Overall, 50\% of males and females expect to live until 86 and 95 years old, respectively, and expected lives increase with savings. 
 
\begin{table}[h!]
 \centering
 \caption{\bf Median Expected Life, by Savings Quintile}\label{table:mediantime}
 \scalebox{1}{ \begin{tabular}{c|ccc}
\toprule
 {\bf Savings} & {\bf Male} & {\bf Female} & {\bf Overall} \\
 \midrule
Q1&85.15&93.80&86.89 \\
 &\small (5.79)&\small (6.03)&\small (5.82) \\
Q2&85.86&94.24&87.64 \\
 &\small (5.81)&\small (6.06)&\small (5.84) \\
Q3&86.45&94.83&88.23 \\
&\small (5.83)&\small (6.09)&\small (5.88) \\
Q4&87.62&95.48&89.40 \\
 &\small (5.88)&\small (6.12)&\small (5.95) \\
Q5&90.87&97.25&93.52 \\
&\small (6.01)&\small (6.21)&\small (6.11) \\
 \hline
{\bf Total } &86.75&94.91&89.57 \\
 &\small (5.82)&\small (6.09)&\small (5.94) \\
 \bottomrule
 \end{tabular}}%
 \caption*{\footnotesize {\bf Note:} The table shows the predicted median expected life at the time of retirement implied by our estimates of the Gompertz mortality distribution. Standard errors are reported in the parentheses.}
\end{table}%

\setcounter{section}{0}
\setcounter{equation}{0}
\setcounter{table}{0}
\setcounter{figure}{0}
\renewcommand{\thesection}{ D}
\renewcommand{\theequation}{D.\arabic{equation}}
\renewcommand{\thetable}{D.\arabic{table}}
\renewcommand{\thefigure}{D.\arabic{figure}}

\section{Determining the Runner-Up Firm\label{section:runnerup}}
We define the runner-up firm in round one as the firm with the highest probability of being chosen in the first round once we exclude the chosen firm.
Furthermore, under the assumption that the runner-up in round one is one of the two most competitive firms in the second round, we can identify the runner-up firm for the second round. 

To construct a measure of the probability of being selected in the first round, we estimate a series of \emph{alternative-specific conditional logit model}. 
To allow for the most general estimation, we divided the sample into 90 different groups, based on the age at retirement (below, at, and above the NRA), gender, channel (recall that we combine insurance companies and sales agents into one, so there are three channels) and balance quintiles. For each group, we estimate the model where the choice of an individual depends on firms' characteristics: e.g., the ratio of reserves to assets, the fraction of sellers employed by each firm, the ratio between the fraction of complaints and premium of each firm, and the risk rating and also the \texttt{mwr}. The following expression gives the random utility associated with $j$'s offer to $i$
\begin{eqnarray}
\eta_{ijt}=\gamma_j^0+\gamma^1\times Z_{jt}+\gamma^2\times \texttt{mwr}_{ij}+ \varepsilon_{ijt},\label{eq:runnerup2}
\end{eqnarray}
where $\gamma_j^0$ is a company-specific constant, and $\gamma^1$ is a coefficient vector for firm-specific variables.
Then the probability of observing a particular choice is then given by $
\Pr(D_{i}^1=j)=\frac{\exp(\hat{\eta}_{ijt})}{\sum_{j=1}^{J}\exp(\hat{\eta}_{ij})}$. 
Using these estimated probabilities for a retiree $i$, we say that a company $j$ is the runner-up if $j$ provides the highest utility to individual $i$ among the set of companies ultimately not chosen by $i$. 

\setcounter{section}{0}
\setcounter{equation}{0}
\setcounter{table}{0}
\setcounter{figure}{0}
\renewcommand{\thesection}{ E}
\renewcommand{\theequation}{E.\arabic{equation}}
\renewcommand{\thetable}{E.\arabic{table}}
\renewcommand{\thefigure}{E.\arabic{figure}}

\section{ Recovering Pension from Expected Present Value}
In this section, we consider the reverse problem of determining pension $P$ from $\rho$ and $b$ for a retiree with bequest preference $\theta$. 
This exercise is important because, if we can uniquely determine pension from the expected present value, then it will allow us to go back and forth between the monetary value of an annuity (for the supply side) to utility for the retiree (for the demand side). From (\ref{eq:utility11}) we know that $w(P, B; \theta)=\rho(P) + \theta b(P)$, and letting $\varpi =w(P, B; \theta)$ we get 
\begin{eqnarray*}
\varpi &=& u(P) \times D^R + u(2\times P) \times D^{R,DP} + \theta \left(v(P) \times D^S + v(0.6\times P) \times D^{S,GP}\right)\\
&=&\frac{P^{-2}}{-2} \left( D^R + \frac{D^{R,DP}}{4} + \theta \left(D^S + \frac{D^{S,GP}}{0.36}\right)\right),
\end{eqnarray*}
where the second equality follows from $u(c) = v(c) = \frac{c^{-2}}{-2}$. 
Then we can solve for the pension  
\begin{eqnarray}
P=\sqrt{\frac{\left(D^R + \frac{D^{R,DP}}{4}\right)+\theta \left(D^S + \frac{D^{S,GP}}{0.36}\right)}{-2\times \varpi}}.\label{eq:Pension}
\end{eqnarray}

\setcounter{section}{0}
\setcounter{equation}{0}
\setcounter{table}{0}
\setcounter{figure}{0}
\renewcommand{\thesection}{ F}
\renewcommand{\theequation}{F.\arabic{equation}}
\renewcommand{\thetable}{F.\arabic{table}}
\renewcommand{\thefigure}{F.\arabic{figure}}

\section{Proofs \label{proofs}}

{\bf Proof of Lemma \ref{lemma1}.}

\begin{proof} 
Note first that, given the proposed strategies, as $\varepsilon$ goes to zero, the winner is the firm with the maximum $\rho_i({P}^{\max}_{ij})+\theta_i\times b_i({P}^{\max}_{ij})+{\beta_{i}\times{Z}_{j}}$.
We introduce some notation and then check that the proposed strategies are optimal for any $\varepsilon >0$:
\begin{itemize}
\item Given a history $\mathfrak {H}$, let $\bf{\tilde{\tilde P}}_{i}$ be the vector of standing offers.

\item Given a history $\mathfrak {H}$ at which $j$ plays, let ${\mathcal E}_1$ be the event that $j= \arg\max_{j \in J}\Big\{ \rho_i({P}^{\max}_{ij})+\theta_i\times b_i({P}^{\max}_{ij})+{\beta_{i}\times{Z}_{j}}\Big\}$; and let $\mu_j(\mathfrak {H}) \equiv \Pr({\mathcal E}_1)$. 

\item Given a history $\mathfrak {H}$ at which $j$ plays and player $k$ is winning (it could be the case that $j=k$), let ${\mathcal E}_2$ be the event that $\tilde{\tilde P}_{il}+\varepsilon>{P}^{\max}_{il}$ for all $l\neq j$ and $l\neq k$. Let $\tilde{\mu}_j(\mathfrak {H})\equiv \Pr({\mathcal E}_2)$ and $\tilde{\tilde{\mu}}_j(\mathfrak {H}) \equiv \Pr({\mathcal E}_1 \wedge {\mathcal E}_2)$.

\item Given $\mathfrak {H}$ and conditional on ${\mathcal E}_1$, define ${P}_{ji}^*$ as the expected value of $P$ such that  
\begin{eqnarray*}
\beta_{i}\times{Z}_{j_i^*}+\theta_i \times b_i(P)+ \rho_i(P)= \max_{k\neq j}\Big\{\beta_{i}\times{Z}_{k} +\theta_i\times b_i({P}^{\max}_{ik})+ \rho_i({P}^{\max}_{ik})\Big\}.
\end{eqnarray*}
Note that ${P}_{ji}^* \leq {P}^{\max}_{ij}$.

\item Given $\mathfrak {H}$ and conditional on ${\mathcal E}_1 \wedge \lnot {\mathcal E}_2$, define $\tilde{P}_{ji}^*$ as the expected value of $P$ such that  
\begin{eqnarray*}
\beta_{i}\times{Z}_{j_i^*}+\theta_i\times b_i(P)+ \rho_i(P)= \max_{k\neq j}\Big\{\beta_{i}\times{Z}_{k} +\theta_i\times b_i({P}^{\max}_{ik})+ \rho_i({P}^{\max}_{ik})\Big\}.
\end{eqnarray*} 
\end{itemize}
Assume first $\mathfrak {H}$ is such that $j$ is not the current winner, then $j$'s expected payment from choosing $Improve$ is greater than the one from choosing $Stay$:
\begin{eqnarray*}
\mu_j(\mathfrak {H}) \times (S_i - UNC_j\times {P}_{ji}^*) \geq (1-\tilde{\mu}_j(\mathfrak {H})) \times \mu_j(\mathfrak {H}) \times (S_i - UNC_j\times {P}_{ji}^*).
\end{eqnarray*}
Assume ($\mathfrak {H}$) is such that $j$ is the current winner. Then $j$'s expected payoff of choosing $Stay$ is
\begin{eqnarray*}
\mu_j(\mathfrak {H}) \times (S_i - UNC_j\times {P}_{ji}^*) = (\mu_j(\mathfrak {H})-\tilde{\tilde{\mu}}_j(\mathfrak {H})) \times (S_i - UNC_j\times \tilde{P}_{ji}^*)+\tilde{\tilde{\mu}}_j(\mathfrak {H}) \times (S_i - UNC_j\times \tilde{\tilde{P}}_{ji}^*), 
\end{eqnarray*}
which is greater than or equal the expected payment of choosing $Improve$, so that 
\begin{eqnarray*}
(\mu_j(\mathfrak {H})-\tilde{\tilde{\mu}}_j(\mathfrak {H})) \times (S_i - UNC_j\times \tilde{P}_{ji}^*)+\tilde{\tilde{\mu}}_j(\mathfrak {H}) \times (S_i - UNC_j \times (\tilde{\tilde{P}}_{ji}^*+\varepsilon)). 
\end{eqnarray*}
\end{proof}

{\bf Proof of Lemma \ref{lemma:Wr}.}
\begin{proof} For notational simplicity, we denote the LHS of Equation (\ref{eq:OLS}) as $\mathfrak{U}$ and the RHS as a sum $\tilde{\beta} +\varpi$, and suppressing the conditioning on savings $S$. Suppose there are $J$ firms.
From the observed chosen pensions and $F_{\theta}$, we can identify the distribution of $\mathfrak{U}$, which is, by definition, also the distribution of the second-highest value of the sum $\tilde{\beta} +\varpi$. We denote the latter distribution by $F_{\tilde{\beta} +\varpi}^{(J-1:J)}(\cdot)$. There is a one-to-one mapping between the distribution of order-statistics and the ``parent" distribution of the sum $F_{\tilde{\beta} +\varpi}(\cdot)$, which is pinned down by $ 
F_{\mathfrak{U}}(t)=F_{\tilde{\beta} +\varpi}^{(J-1:J)}(t) = J (J-1) \int_{0}^{F_{\tilde{\beta} +\varpi}(t)} (\xi^{J-2}\times \xi )d\xi.
$
 
Then, using the fact that $F_{\tilde{\beta} +\varpi}= F_{\tilde{\beta}} * F_{\varpi}$, is a convolution, where $*$ is the convolution operator, we can identify the distribution of $\varpi$ using deconvolution method. Lastly, we observe that there is a one-to-one mapping from $\varpi$ to ${P}^{\max}$ --the maximum pension runner-up firm can offer to retiree (see Equation \ref{eq:Pension}), which we denote by a function ${P}^{\max}= m(\varpi)=S/UNC_k$. 
Then we get  
\begin{eqnarray*}
W_r(\xi) &=& \Pr(r\leq \xi) = \Pr\left(\frac{UNC_k}{UNC_i}\leq \xi\right)=\Pr\left(\frac{S}{{P}^{\max}} \leq \xi \times UNC_i\right)\\
&=& \Pr\left({P}^{\max}\geq \frac{S}{\xi\times UNC_i}\right)=1-\Pr\left({P}^{\max}\leq \frac{S}{\xi\times UNC_i}\right)\\
&=&1-\Pr\left(m^{-1}({P}^{\max})\leq m^{-1}\left(\frac{S}{\xi\times UNC_i}\right)\right)=1-\Pr\left(\varpi\leq m^{-1}\left(\frac{S}{\xi\times UNC_i}\right)\right)\\
&=& 1-F_{\varpi}\left(m^{-1}\left(\frac{S}{\xi\times UNC_i}\right)\right), \qquad (\because {P}^{\max} = S/UNC_k).
\end{eqnarray*}
\end{proof}}

\end{document}